\documentclass[pra,aps,preprint,amsmath,amssymb,showkeys]{revtex4}
\usepackage{lineno,hyperref}
\usepackage{graphicx}
\usepackage{epsfig}
\usepackage{epstopdf}

\include{graphics}
\begin{document}

%\begin{frontmatter}

\title{Strong effects of fast collisions between pulsed optical beams in a 
linear medium with weak cubic loss}

\author{Avner Peleg$^{1}$ and Toan T. Huynh$^{2}$}

\affiliation{$^{1}$Department of Mathematics, Azrieli College of Engineering, Jerusalem 9371207, Israel}

\affiliation{$^{2}$Department of Mathematics, University of Medicine and Pharmacy at Ho Chi Minh City, Ho Chi Minh City, Vietnam}

\date{\today}

\begin{abstract}
We investigate fast collisions between pulsed optical beams in a linear medium 
with weak cubic loss that arises due to nondegenerate two-photon absorption. 
We introduce a perturbation method with two small parameters and use it to 
obtain general formulas for the collision-induced changes in the pulsed-beam's 
shape and amplitude. Moreover, we use the method to design and characterize collision 
setups that lead to strong localized and nonlocalized intensity reduction effects. 
The values of the collision-induced changes in the pulsed-beam's shape in 
both setups are larger by one to two orders of magnitude compared with 
the values obtained in previous studies of fast two-pulse collisions.  
Furthermore, we show that for nonlocalized setups, the graph of the 
collision-induced amplitude shift vs the difference between the 
first-order dispersion coefficients for the two pulsed-beams 
has two local minima. This finding represents the first observation 
of a deviation of the graph from the common funnel shape that was obtained 
in all previous studies of fast two-pulse collisions in the presence of 
weak nonlinear loss. The predictions of our perturbation theory are in 
good agreement with results of numerical simulations with the perturbed 
linear propagation model, despite the strong collision-induced effects. 
Our results can be useful for multisequence optical communication links 
and for reshaping of pulsed optical beams. 
\end{abstract}

\keywords{Pulsed optical beams, two-photon absorption, beam collisions}

\maketitle
%\newpage

%\end{frontmatter}

\section{Introduction}
\label{Introduction}

Collisions between optical pulses play an important role 
in many optical systems. Examples include 
wavelength-division-multiplexed (WDM) optical communication links \cite{Agrawal2020,Tkach97,Mollenauer2006,Essiambre2010,Iannone98}, 
fiber grating \cite{Agrawal2020,Stegeman90,Broderick97}, 
waveguide couplers \cite{Agrawal2020,Haus91,Huang94}, 
and Sagnac interferometers \cite{Agrawal2020,Betts92,Mori95}. 
On one hand, the collisions can be beneficially used for optical 
switching \cite{Agrawal2020,Broderick97,Agrawal2019,Moores91,Rothenberg93}, 
pulse compression \cite{Agrawal2019,Agrawal89,Rothenberg90}, 
generation of pulse trains \cite{Betts92}, 
parametric amplification \cite{Mori95}, 
and optical logical gates \cite{Stieglitz2000}. On the other hand, the collisions 
can have harmful effects in WDM communication systems. In these systems, many pulse 
sequences propagate through the same optical medium. The pulses in each sequence  
propagate with the same group velocity, but the group velocities are different 
for pulses from different sequences. As a result, collisions between pulses from 
different sequences are very frequent, and their cumulative effect can lead to 
severe transmission degradation and to transmission error 
\cite{Agrawal2020,Tkach97,Mollenauer2006,Essiambre2010,Iannone98}. 
For these reasons, many research efforts have been devoted to 
studying the different effects of collisions between optical 
pulses \cite{Agrawal2020,Tkach97,Mollenauer2006,Essiambre2010,Iannone98,Agrawal2019}.

In the current work, we study fast collisions between two pulsed 
optical beams in a linear bulk optical medium with weak cubic loss. 
We assume that the cubic loss is due to nondegenerate two-photon 
absorption (2PA), i.e., due to simultaneous absorption of two photons 
with different wavelengths \cite{Hagan2002,Hagan2011}. Fast two-pulse (or two-beam) 
collisions are encounters between two optical pulses (or optical beams), 
in which the interval along which the two pulses overlap (the collision length)
is much smaller than all the other length scales in the problem 
\cite{fast_collisions}. As described in the preceding paragraph, 
collisions between optical pulses are very important in WDM communication 
links. The nonlinearities in these systems are typically weak 
\cite{Tkach97,Agrawal97}, and almost all collisions are fast 
\cite{PNH2017,Mollenauer2003,Nakazawa2000}. Therefore, the study 
of fast collisions of optical pulses or optical beams in the presence 
of weak nonlinearities is very relevant to WDM optical communication links.   
An important problem in this area concerns the characterization 
of the effects of a single fast two-pulse collision by explicit formulas. 
Indeed, such characterization enables the evaluation of the cumulative effects 
of many fast collisions without having to perform a large number of 
numerical simulations or experiments. Furthermore, it enables the design 
of methods for compensating the harmful cumulative effects of fast 
optical pulse collisions.

In Refs. \cite{PNH2017,NHP2022}, we studied the dynamics of fast two-pulse 
collisions in linear optical waveguides with weak cubic loss. We first developed 
a perturbation method for analyzing the effects of these collisions. 
We then used the method to show that the expression for the collision-induced 
amplitude shift in a single collision has the same simple form as the expression for the amplitude 
shift in a fast collision between two solitons of the cubic nonlinear Schr\"odinger 
equation in the presence of weak cubic loss \cite{optical_solitons}. 
We also found that within the leading order of the perturbation theory, 
the shapes of the optical pulses are not changed by the collision. 
Furthermore, we showed that similar behavior holds in fast 
collisions between two concentration pulses in the presence of weak quadratic loss 
in systems described by weakly perturbed diffusion-advection equations 
\cite{PNH2017,NHP2022}. The predictions of our perturbation theory were 
confirmed by numerical simulations with the perturbed linear propagation 
equation and the perturbed linear diffusion-advection equation with weak 
nonlinear loss. The results of Refs. \cite{PNH2017,NHP2022} are quite surprising. 
Indeed, the pulses in weakly perturbed linear optical waveguides are not shape 
preserving. As a result, the common expectation is that conclusions drawn from analysis       
of optical soliton collisions would {\it not} be applicable for collisions between optical pulses 
in weakly perturbed linear waveguides \cite{Tkach97,Agrawal2019,Agrawal89,PCG2003}. 
However, in Refs. \cite{PNH2017,NHP2022}, we showed that exactly the opposite is true.

In Ref. \cite{PHN2022}, we generalized the perturbation method of Refs. \cite{PNH2017,NHP2022} 
to treat fast collisions between two time-independent optical beams in the presence of weak 
cubic loss in spatial dimension 2. In this case, the collisions are induced by beam-steering, 
and special techniques are needed to realize and control the steering 
\cite{McManamon2009,Brandl2013,Oh2016}. We used the generalized perturbation 
approach to show that a fast collision between two time-independent beams leads 
to a change in the beam shapes in the direction transverse to the vector of relative 
velocity between the beams. Additionally, we studied the case of a separable initial condition 
for both beams, which is of special importance, since it describes the electric fields that 
are produced by many types of lasers \cite{Siegman86,Kogelnik66}. In this case we found that 
the beam shape in the longitudinal direction is not changed by the collision within the leading 
order of the perturbative calculation. Furthermore, we showed that for a separable initial 
condition, the longitudinal part in the expression for the amplitude shift is universal, 
while the transverse part is not universal and is proportional to the integral of 
the product of the beam intensities with respect to the transverse coordinate. 
The predictions of the generalized perturbation theory were confirmed by numerical 
simulations with the perturbed linear propagation equation with weak cubic loss.

The studies in Refs. \cite{PNH2017,NHP2022,PHN2022} were limited to weak 
collisional effects, that is, the values of the collision-induced amplitude 
shift and the intensity reduction factor in the collision setups that were 
considered in these works were small. Although we were able to measure these 
values in numerical simulations, it is very difficult to measure such small values 
in experiments. Furthermore, the works in Refs. \cite{PNH2017,NHP2022,PHN2022} 
were limited to spatial dimensions 1 and 2 \cite{spatial_dimension}. 
As a result, an important property of the collisional effects that exists 
only in spatial dimension 3, i.e., in collisions between pulsed optical beams, 
was not addressed in these studies. This property is related to the behavior 
of the graph of the collision-induced amplitude shift vs the difference between 
the beam velocities, which is one of the main tools for analyzing the collisional 
effects. According to the perturbation theory of Refs. \cite{PNH2017,NHP2022,PHN2022}, 
this graph has a funnel shape, and does not possess any local extrema. 
Furthermore, no deviation from the common funnel shape form can exist in the graph 
for fast collisions in spatial dimensions 1 and 2. This property was verified by 
numerical simulations in Refs. \cite{PNH2017,NHP2022,PHN2022}, and was also 
observed in fast collisions between optical solitons in the presence of weak 
nonlinear loss \cite{PNC2010,PC2012,NH2021,NH2022}.       
Another important aspect of the fast collision problem that was not addressed 
in previous works concerns the characterization of the differences between the 
collision-induced effects in localized and nonlocalized collision setups. 
More specifically, as will be shown in the current paper, the perturbation 
theory's formulas for the collision-induced changes in the beam shapes enable 
the design of localized collision setups, in which the intensity reduction is 
concentrated mainly near the beam centers, and nonlocalized collision setups, 
in which the intensity reduction affects the entire main bodies of the beams. 
The intensity reduction patterns observed in these setups are very different, 
and these differences can also strongly affect the magnitude of the 
collision-induced amplitude shift and its dependence on the difference 
between the beam velocities. Despite the importance of this aspect of the 
problem, no attempt to characterize the differences between the collisional 
effects in localized and nonlocalized collision setups was made in previous studies.

In the current paper, we address the important aspects of the fast 
two-beam collision problem that were overlooked in previous studies. 
More specifically, we study fast collisions between two pulsed optical 
beams in a linear optical medium with weak cubic loss in physical setups, 
in which the collisional effects are relatively strong. The collisions 
are induced by the difference between the first-order dispersion 
coefficients for the two pulsed-beams. These collisions are easier to 
realize than the collisions between time-independent beams that were 
studied in Ref. \cite{PHN2022}, since their experimental realization 
does not require the application of beam steering. 
We first introduce a perturbation approach for calculating the 
effects of a single fast collision between two pulsed-beams. 
We then use the approach to obtain general explicit expressions for 
the collision-induced changes in the pulsed-beam's shape and amplitude.
Moreover, we use the approach to design and characterize collision setups 
that lead to strong localized and nonlocalized intensity reduction effects. 
The values of the collision-induced changes in the pulsed-beam's shape 
predicted by our perturbation theory are larger by one to two orders 
of magnitude compared with the values obtained in fast collisions 
between time-independent beams in Ref. \cite{PHN2022} and in fast 
collisions between optical solitons in spatial dimension 1 
in Ref. \cite{PC2020}. The perturbation theory predictions are 
in good agreement with the results of numerical simulations 
with the perturbed linear propagation model for both localized and 
nonlocalized collision setups, despite the relatively strong 
collision-induced effects.

We gain further insight into the effects of the collisions 
by studying the dependence of the following two central physical quantities 
on the difference between the first-order dispersion coefficients for 
the two pulsed-beams. (1) The final value of the intensity reduction ratio 
on the propagation axis (the $z$ axis). (2) The collision-induced amplitude 
shift. The results of the perturbative calculation and numerical simulations 
for the final value of the intensity reduction ratio on the $z$ axis show 
that the significant intensity reduction effects are not limited to 
intermediate differences between the first-order dispersion coefficients, 
but also exist for large differences. These results hold for both localized 
and nonlocalized collision setups. Furthermore, the results of 
the numerical simulations for the nonlocalized collision setups show that 
the graph of the amplitude shift vs the difference between the first-order 
dispersion coefficients has two local minima at intermediate values of 
the difference. These local minima are correctly captured by our perturbative 
calculation. To our knowledge, this finding represents the first observation 
of a deviation of the graph of the amplitude shift vs the difference between 
the first-order dispersion coefficients from the common funnel shape that is 
obtained in fast collisions between temporal pulses or time-independent beams 
in linear optical media \cite{PNH2017,NHP2022,PHN2022}, and in fast collisions 
between optical solitons \cite{PNC2010,PC2012,NH2021,NH2022}.

We focus our attention on two-beam collisions in the presence of 
cubic loss, since cubic loss is important in many optical 
systems \cite{Agrawal2007,Dekker2007,Borghi2017,Boyd2008}. 
Furthermore, cubic loss is the dominant nonlinear loss 
mechanism in these systems, and is therefore important 
in fast optical pulse collisions \cite{PNH2017,NHP2022,PHN2022,PNC2010}. 
The optical medium's cubic loss is typically due to 2PA 
\cite{Agrawal2007,Dekker2007,Borghi2017,Boyd2008}. 
Propagation of optical pulses and optical beams in the presence 
of cubic loss has been studied in many earlier works, both in weakly 
perturbed linear media \cite{PNH2017,NHP2022,Perry97,Liang2005,Cohen2005b,Cohen2004}, 
and in nonlinear media \cite{PNC2010,PC2020,Malomed89,Stegeman89,Aceves92,Tsoy2001,Gaeta2012,PC2018}.
The subject gained further attention in recent years due to the importance of 2PA 
in silicon nanowaveguides, which are expected to play a key role in many applications 
in optoelectronic devices \cite{Agrawal2007,Dekker2007,Borghi2017,Gaeta2008}.       
In the current paper, we study the effects of weak cubic loss due to 
nondegenerate 2PA on collisions in a bulk optical medium. We neglect the 
effects of degenerate 2PA, which arises due to the simultaneous absorption 
of two photons with the same wavelength. This means that we assume that 
the effects of cubic loss on single-beam propagation are much weaker compared 
with the effects of cubic loss on interbeam interaction. This situation can 
be realized, for example, in certain nonlinear semiconductors, 
in which degenerate 2PA is much weaker than nondegenerate 2PA 
\cite{Hagan2002,Hagan2011,Rauscher97}. We also assume that the 
optical medium is weakly nonlinear and neglect the effects 
of cubic (Kerr) nonlinearity. We point out that this assumption 
was successfully used in previous experimental and theoretical works, see, e.g., 
Refs. \cite{Perry97,Liang2005,Cohen2005b,Cohen2004}. For similar reasons, 
we neglect the effects of high-order nonlinear loss on the collisions,  
and remark that the latter effects can be described by the same perturbation 
approach that is presented in the current paper (see also Ref. \cite{QMN2022}, 
where the calculation was carried out for pulse collisions 
in spatial dimension 1).

The rest of the paper is organized in the following manner. In Section 
\ref{theory}, we present our perturbation approach for fast two-beam collisions and 
its predictions for the collision-induced effects. In Section \ref{setups}, 
we describe the guiding principles for the design of collision setups 
that lead to strong localized and nonlocalized intensity reduction. 
We also present the calculation of the collision-induced changes in the 
pulsed-beam's shape and amplitude, and the calculation of the intensity 
reduction ratio for these setups. In Section \ref{simu}, we present the 
results of our numerical simulations with the perturbed linear propagation 
model, and compare these results with the perturbation theory predictions. 
We summarize our conclusions in section \ref{conclusions}.

\section{Theoretical predictions for collision-induced effects}
\label{theory}

\subsection{Introduction}
\label{th_intro}
      
We consider the dynamics of fast collisions between two pulsed optical 
beams in a three-dimensional linear optical medium with weak cubic loss, 
where the cubic loss arises due to nondegenerate 2PA. 
We assume that the pulsed-beams propagate along the $z$ axis and that the 
propagation is accurately described by the paraxial approximation 
\cite{Siegman86,Kogelnik66,Ishimaru2017}. In addition to weak cubic loss, 
we take into account the effects of first-order and second-order dispersion 
and isotropic diffraction. Therefore, the dynamics of the fast collision is 
described by the following weakly perturbed linear propagation model:      
\begin{eqnarray}&&
\!\!\!\!\!\!\!\!\!\!\!\!\!\!\!\!\!\!\!\!
i\partial_{z''}E'_{1} + i\tilde\beta_{11}\partial_{\tau'}E'_{1} 
-\tilde\beta_{21}\partial^{2}_{\tau'} E'_{1}/2 
+ \tilde d_{21}\partial^{2}_{x'} E'_{1} 
+ \tilde d_{21}\partial^{2}_{y'} E'_{1}
= - 2i\rho_{3}|E'_{2}|^{2}E'_{1},
\nonumber\\&&
\!\!\!\!\!\!\!\!\!\!\!\!\!\!\!\!\!\!\!\!
i\partial_{z''}E'_{2} + i\tilde\beta_{12}\partial_{\tau'}E'_{2} 
-\tilde\beta_{22}\partial^{2}_{\tau'} E'_{2}/2 
+ \tilde d_{22}\partial^{2}_{x'} E'_{2} 
+ \tilde d_{22}\partial^{2}_{y'} E'_{2}
= - 2i\rho_{3}|E'_{1}|^{2}E'_{2}.
\label{pb1}
\end{eqnarray}
In Eq. (\ref{pb1}), $E'_{j}$ with $j=1,2$ are the dimensional electric fields 
of pulsed-beams 1 and 2, $z''$ and $\tau'$ are the dimensional propagation 
distance and time, and $x'$ and $y'$ are the dimensional spatial coordinates 
in the $xy$ plane. In addition, $\tilde\beta_{1j}$ and $\tilde\beta_{2j}$ 
are the dimensional first-order and second-order dispersion coefficients, 
$\tilde d_{2j}=\lambda_{j}/(4\pi)$ are the dimensional diffraction coefficients, 
where $\lambda_{j}$ are the wavelengths, and $\rho_{3}$ is the dimensional 
cubic loss coefficient.

We now make a change of variables of the form 
\begin{eqnarray}&&
z'=z'', \;\;\;\; \tau = \tau' - \tilde\beta_{11} z', 
\;\;\;\; E_{1}(z',x',y',\tau)= E'_{1}(z'',x',y',\tau').  
\label{pb2}
\end{eqnarray}
This means that we go to the retarded reference frame for 
pulsed-beam 1 (see, e.g., Ref. \cite{Agrawal2019}, p. 61). 
The transformation (\ref{pb2}) is important, since it enables the 
identification of the true small parameters in the fast collision    
problem, and as a result, the application of the perturbation method 
and the derivation of explicit approximate formulas for the collision-induced 
changes in the shapes and amplitudes of the pulsed-beams. 
Using the transformation (\ref{pb2}) in Eq. (\ref{pb1}), we obtain: 
\begin{eqnarray}&&
\!\!\!\!\!\!\!\!\!\!\!\!\!\!\!\!\!\!\!\!
i\partial_{z'}E_{1} -\tilde\beta_{21}\partial^{2}_{\tau} E_{1}/2 
+ \tilde d_{21}\partial^{2}_{x'} E_{1} 
+ \tilde d_{21}\partial^{2}_{y'} E_{1}
= - 2i\rho_{3}|E_{2}|^{2}E_{1},
\nonumber\\&&
\!\!\!\!\!\!\!\!\!\!\!\!\!\!\!\!\!\!\!\!
i\partial_{z'}E_{2} + i\Delta\tilde\beta_{1}\partial_{\tau}E_{2} 
-\tilde\beta_{22}\partial^{2}_{\tau} E_{2}/2 
+ \tilde d_{22}\partial^{2}_{x'} E_{2} 
+ \tilde d_{22}\partial^{2}_{y'} E_{2}
= - 2i\rho_{3}|E_{1}|^{2}E_{2}, 
\label{pb3}
\end{eqnarray}
where $\Delta\tilde\beta_{1}=\tilde\beta_{12} - \tilde\beta_{11}$.

In order to employ the perturbation theory for the fast collision, 
we must bring the propagation model to a nondimensional 
form \cite{nondimensional}. For this purpose, we define 
the dimensionless propagation distance $z$, the dimensionless time $t$, 
the dimensionless spatial coordinates $x$ and $y$, and the dimensionless  
electric fields $\psi_{j}$ by: 
\begin{eqnarray}&&
\!\!\!\!\!\!\!\!\!\!\!\!\!\!\!
z=z'/(2L_{D}), \;\;\;\; 
t=\tau/\tau_{0}, \;\;\;\;
x=x'/x'_{0}, \;\;\;\; 
y=y'/x'_{0}, \;\;\;\;  
\psi_{j}=E_{j}/\sqrt{P_{0}}.  
\label{pb4}
\end{eqnarray}  
In Eq. (\ref{pb4}), $L_{D}=\tau_{0}^{2}/|\tilde\beta_{21}|$ is 
the dispersion length, $\tau_{0}$ is the temporal width of a 
reference pulsed-beam, $x'_{0}$ is the width of a reference 
pulsed-beam along the $x$ axis, and $P_{0}$ is the peak power 
of the reference pulsed-beam. Using the relations (\ref{pb4}) 
in Eq. (\ref{pb3}), we obtain the dimensionless form of the 
weakly perturbed propagation model: 
\begin{eqnarray}&&
\!\!\!\!\!\!\!\!\!\!\!\!\!\!\!\!\!\!
i\partial_{z}\psi_{1} - \mbox{sgn}(\tilde\beta_{21}) \partial^{2}_{t} \psi_{1} 
+ d_{21}\partial^{2}_{x} \psi_{1} + d_{21}\partial^{2}_{y} \psi_{1}
= -2i\epsilon_{3}|\psi_{2}|^{2}\psi_{1},
\nonumber\\&&
\!\!\!\!\!\!\!\!\!\!\!\!\!\!\!\!\!\!
i\partial_{z}\psi_{2} + i\Delta\beta_{1}\partial_{t}\psi_{2}
+ \beta_{22}\partial^{2}_{t} \psi_{2} + d_{22}\partial^{2}_{x} \psi_{2}
+  d_{22}\partial^{2}_{y} \psi_{2}
= -2i\epsilon_{3}|\psi_{1}|^{2}\psi_{2}. 
\label{pb5}
\end{eqnarray}   
The coefficients $\Delta\beta_{1}$ and $\beta_{22}$ in Eq. (\ref{pb5}) 
are the dimensionless first-order and second-order dispersion coefficients, 
$d_{2j}$ are the dimensionless diffraction coefficients, and $\epsilon_{3}$ is the 
dimensionless cubic loss coefficient. These coefficients are defined by 
the following relations: 
\begin{eqnarray}&&
\!\!\!\!\!\!\!\!\!\!\!\!\!\!\!
\Delta\beta_{1}= 2\Delta\tilde\beta_{1}\tau_{0}/|\tilde\beta_{21}|, \;\;\;
\beta_{22}= -\tilde\beta_{22}/|\tilde\beta_{21}|, \;\;\;
d_{2j}= \tau_{0}^{2}\lambda_{j}/(2\pi |\tilde \beta_{21}|x^{\prime 2}_{0}),  
\nonumber\\&&
\epsilon_{3}=2P_{0}\tau_{0}^{2}\rho_{3}/|\tilde\beta_{21}|. 
\label{pb6}
\end{eqnarray} 
In the current paper, we study fast collisions in the presence of weak 
cubic loss. We therefore assume that the coefficients $\Delta\beta_{1}$ 
and $\epsilon_{3}$ satisfy $|\Delta\beta_{1}| \gg 1$ and $0 < \epsilon_{3} \ll 1$.

Our perturbation approach applies for fast collisions between 
pulsed optical beams with general initial shapes, such that the total 
energies $\int_{-\infty}^{\infty} dt  \int_{-\infty}^{\infty} dx 
\int_{-\infty}^{\infty} dy \, |\psi_{j}(t,x,y,0)|^{2}$ are finite. 
We assume that the initial pulsed-beams can be characterized by the 
following parameters. (1) The initial amplitudes $A_{j}(0)$. 
(2) The initial widths of the pulsed-beams along the $t$, $x$, and $y$ axes, 
$W_{j0}^{(t)}$, $W_{j0}^{(x)}$, and $W_{j0}^{(y)}$. 
(3) The initial positions of the beam centers $(t_{j0}, x_{j0}, y_{j0})$.   
(4) The initial phases $\alpha_{j0}$. Thus, the initial electric fields 
can be written as: 
\begin{eqnarray} &&
\psi_{j}(t,x,y,0)=A_{j}(0)h_{j}(t,x,y)\exp(i\alpha_{j0}),  
\label{pb7}
\end{eqnarray} 
where $h_{j}(t,x,y)$ is a real-valued function that characterizes the 
initial spatio-temporal distribution of the electric field.
We are equally interested in the important case, where the initial 
electric fields of the two pulsed-beams are separable, i.e., where 
each of the functions $\psi_{j}(t,x,y,0)$ is a product of three 
functions of $t$, $x$, and $y$:  
\begin{eqnarray} &&
\psi_{j}(t,x,y,0)=A_{j}(0) h_{j}^{(t)}[(t-t_{j0})/W_{j0}^{(t)}]
h_{j}^{(x)}[(x-x_{j0})/W_{j0}^{(x)}]
\nonumber \\&&
\times 
h_{j}^{(y)}[(y-y_{j0})/W_{j0}^{(y)}]\exp(i\alpha_{j0}).  
\label{pb8}
\end{eqnarray}   
This initial condition is of special importance since it describes 
the electric fields that are produced by many lasers \cite{Siegman86,Kogelnik66}.

In the current paper, we demonstrate strong collision-induced effects in 
complete fast collisions. We therefore obtain conditions on the 
physical parameter values for these collisions. The complete collision assumption 
means that the pulsed-beams are well-separated before and after the collision. 
As a result, in these collisions, the values of the $t$ coordinate of the 
pulsed-beam centers at $z=0$ and at the final propagation distance $z_{f}$, 
$t_{j0}$ and $t_{j}(z_f)$, must satisfy $|t_{20}-t_{10}| \gg W_{10}^{(t)}+W_{20}^{(t)}$ 
and $|t_{2}(z_f)-t_{1}(z_f)| \gg W_{1}^{(t)}(z_f)+W_{2}^{(t)}(z_f)$, 
where $W_{j}^{(t)}(z_f)$ are the pulsed-beam widths along the $t$ axis at $z=z_{f}$.         
To obtain the condition for a fast collision, we define the collision length 
$\Delta z_{c}$ as the distance along which the temporal widths of the pulsed-beams  
overlap. From this definition it follows that 
$\Delta z_{c}=2(W_{10}^{(t)}+W_{20}^{(t)})/|\Delta\beta_1|$. 
The fast collision assumption means that $\Delta z_{c}$ is 
much smaller than the length scale $z_{D}^{(min)}$, which is  
the smallest dispersion length or diffraction length in the problem.  
By definition, $z_{D}^{(min)}=\min \left\{ z_{d1}^{(t)}, z_{d2}^{(t)}, 
z_{D1}^{(x)}, z_{D2}^{(x)}, z_{D1}^{(y)}, z_{D2}^{(y)}\right\}$, 
where $z_{dj}^{(t)}$ are the dispersion lengths of the pulsed-beams, 
and $z_{Dj}^{(x)}$ and $z_{Dj}^{(y)}$ are the diffraction lengths 
along the $x$ and $y$ axes. Requiring that $\Delta z_{c} \ll z_{D}^{(min)}$, 
we obtain $2(W_{10}^{(t)}+W_{20}^{(t)}) \ll |\Delta\beta_1| z_{D}^{(min)}$, as 
the condition for a fast collision.

\subsection{The perturbation approach and its predictions for a general initial condition}
\label{th_general}
      
We present here a relatively brief description of the 
perturbative calculation of the collision-induced changes 
in the shapes and amplitudes of the pulsed-beams for the 
general initial condition (\ref{pb7}). 
The results for the separable initial condition (\ref{pb8}) 
are presented in section \ref{th_separable}.    
The current treatment is a generalization of the calculation that was carried out 
in Ref. \cite{PHN2022} for fast collisions between time-independent 
optical beams (in the absence of dispersion effects). 
Since the main steps in the current perturbative calculation are 
similar to the ones described in Ref. \cite{PHN2022} for the time-independent case, 
we concentrate here only on those steps and results, which are essential for 
the understanding of the material in sections \ref{setups} and \ref{simu}. 
We refer the reader, who is interested in more details, to the description 
of the time-independent version of the approach in Ref. \cite{PHN2022}.

In the first step in the perturbative calculation, we look 
for a solution of Eq. (\ref{pb5}) in the form: 
\begin{eqnarray}&&
\psi_{j}(t,x,y,z)=\psi_{j0}(t,x,y,z)+\phi_{j}(t,x,y,z), 
\label{pb11}
\end{eqnarray}           
where $j=1,2$, $\psi_{j0}$ are the solutions of the unperturbed 
linear propagation equations, and $\phi_{j}$ describe corrections to 
the $\psi_{j0}$ due to the effects of cubic loss on the collision. 
By definition, the $\psi_{j0}$ satisfy the equations
\begin{eqnarray}&&
i\partial_{z}\psi_{10} - \mbox{sgn}(\tilde\beta_{21}) \partial^{2}_{t} \psi_{10} 
+ d_{21}\partial^{2}_{x} \psi_{10} + d_{21}\partial^{2}_{y} \psi_{10} = 0,
\label{pb12}
\end{eqnarray}         
and
\begin{eqnarray}&&
i\partial_{z}\psi_{20} + i\Delta\beta_{1}\partial_{t}\psi_{20}
+ \beta_{22}\partial^{2}_{t} \psi_{20} + d_{22}\partial^{2}_{x} \psi_{20}
+  d_{22}\partial^{2}_{y} \psi_{20} = 0.  
\label{pb13}
\end{eqnarray}               
We expand the $\phi_{j}$ in perturbation series with respect to 
the two small parameters $\epsilon_{3}$ and $1/|\Delta\beta_{1}|$. 
We are interested in the first nonzero term in each of these expansions. 
These first nonzero terms in the expansions represent the 
leading-order collision-induced changes in the pulse shapes, 
and we therefore refer to them as the leading-order expressions 
for the $\phi_{j}$.

We substitute the relation (\ref{pb11}) into Eq. (\ref{pb5}) 
and use Eqs. (\ref{pb12}) and (\ref{pb13}) to obtain equations 
for the dynamics of the $\phi_{j}$. We concentrate on the calculation 
of $\phi_{1}$, since the calculation of $\phi_{2}$ is similar. 
To obtain the equation for the leading-order expression for $\phi_{1}$ 
we must neglect high-order terms containing products of $\epsilon_{3}$ 
with $\phi_{1}$ or $\phi_{2}$ in the equation obtained after the 
substitution. Thus, the substitution and the subsequent approximations 
yield the following equation for the leading-order expression for $\phi_{1}$:  
\begin{eqnarray}&&
i\partial_z\phi_{1} - \mbox{sgn}(\tilde\beta_{21}) \partial^{2}_{t} \phi_{1} 
+ d_{21}\partial_{x}^{2}\phi_{1} + d_{21}\partial_{y}^{2}\phi_{1}=
-2i\epsilon_{3}|\psi_{20}|^2\psi_{10}. 
\label{pb14}
\end{eqnarray}
Note that for brevity and simplicity of notation, in Eq. (\ref{pb14}), 
we denote the leading-order expression for $\phi_{1}$ by $\phi_{1}$. 
This notation is also used in the remainder of the paper.

In solving the equation for $\phi_{1}$, we recognize two different 
intervals along the $z$ axis, the collision interval and the 
post-collision interval. To define these intervals, we introduce  
the collision distance $z_{c}$, which is the distance at which the 
$t$ coordinates of the pulsed-beam centers coincide, i.e., 
$t_{1}(z_{c})=t_{2}(z_{c})$. The collision interval is the small 
interval $z_{c} - \Delta z_{c}/2 \le z \le z_{c} + \Delta z_{c}/2$  
around $z_{c}$, in which the two pulsed-beams overlap. This interval 
is an effective boundary layer in our perturbative calculation. 
The post-collision interval is the interval $z > z_{c} + \Delta z_{c}/2$, 
in which the pulsed-beams no longer overlap.

\subsubsection{Calculation of the collision-induced effects in the collision interval}    

We substitute $\psi_{j0}(t,x,y,z)=A_{j}(z)\Psi_{j0}(t,x,y,z)\exp[i\chi_{j0}(t,x,y,z)]$ and 
$\phi_{1}(t,x,y,z)=\Phi_{1}(t,x,y,z)\exp[i\chi_{10}(t,x,y,z)]$ into Eq. (\ref{pb14}),  
where $A_{j}(z)$ are the $z$-dependent amplitudes of the pulsed-beams, 
and $\Psi_{j0}$ and $\chi_{j0}$ are real-valued. This substitution yields 
an equation for $\Phi_{1}$ (see also Ref. \cite{PHN2022} for the time-independent case). 
Neglecting the high-order terms in the latter equation, we arrive at 
the following equation for $\Phi_{1}$ in the leading order of the 
perturbative calculation: 
\begin{eqnarray} &&
\partial_{z}\Phi_{1}= -2\epsilon_{3}A_{1}(z)A_{2}^{2}(z)
\Psi_{20}^{2}\Psi_{10}.
\label{pb15}
\end{eqnarray} 
As we will see, the simple form of the equation for the dynamics 
of $\Phi_{1}$ in the collision interval is important in enabling 
the design of collision setups that lead to strong intensity 
reduction effects.

We calculate the collision-induced amplitude shift of pulsed-beam 1 
from the net change in $\Phi_{1}$ in the collision interval, 
$\Delta\Phi_{1}(t,x,y,z_{c})=\Phi_{1}(t,x,y,z_{c}+\Delta z_{c}/2)-
\Phi_{1}(t,x,y,z_{c}-\Delta z_{c}/2)$. $\Delta\Phi_{1}(t,x,y,z_{c})$ 
is calculated by integration of Eq. (\ref{pb15}) with respect to $z$  
over the collision interval: 
\begin{eqnarray}&&
\!\!\!\!\!\!\!\!\!\!\!\!
\Delta\Phi_{1}(t,x,y,z_{c})\!=\!-2\epsilon_{3}
\!\!\int_{z_{c}-\Delta z_{c}/2}^{z_{c}+\Delta z_{c}/2} 
\!\!\!\!  dz' \, A_{1}(z') A_{2}^{2}(z')
\Psi_{10}(t,x,y,z')\Psi_{20}^{2}(t,x,y,z') . 
%\nonumber \\&&
\label{pb16}
\end{eqnarray}  
We notice that $\Psi_{20}$ is the only function in the integrand on 
the right hand side of Eq. (\ref{pb16}) that contains fast variations 
in $z$, which are of order 1. Therefore, we can approximate 
$A_{j}(z)$ and $\Psi_{10}(t,x,y,z)$ by $A_{j}(z_{c}^{-})$ 
and $\Psi_{10}(t,x,y,z_{c})$, where $A_{j}(z_{c}^{-})$ 
is the limit from the left of $A_{j}(z)$ at $z_{c}$.      
Additionally, since outside of the collision interval loss 
is negligible, we can set $A_{j}(z_{c}^{-})=A_{j}(0)$.  
Furthermore, in calculating the integral, we can take into account 
in an exact manner only the fast dependence of $\Psi_{20}$ on $z$, 
which is contained in the factors $\tilde t=t-t_{20}-\Delta\beta_{1}z$, and replace 
$z$ by $z_{c}$ everywhere else in the expression for $\Psi_{20}$.
This approximation of $\Psi_{20}(t,x,y,z)$ is denoted by 
$\bar\Psi_{20}(\tilde t,x,y,z_{c})$. In addition, we assume that 
the approximate integrand $\bar\Psi_{20}^{2}(\tilde t,x,y,z_{c})$   
is sharply peaked in a small interval around $z_{c}$. As a result, 
we can extend the integral's limits to $-\infty$ and $\infty$ 
(see also Ref. \cite{PHN2022}). Carrying out all these approximations 
and also changing the integration variable from $z'$ to 
$\tilde t=t-t_{20}-\Delta\beta_{1}z'$, we obtain: 
\begin{eqnarray} &&
\!\!\!\!\!\!\!\!\!\!\!\!\!\!\!
\!\!\!\!\!\!\!\!\!\!\!\!\!
\Delta\Phi_{1}(t,x,y,z_{c})\!=\!-\frac{2\epsilon_{3}
A_{1}(0) A_{2}^{2}(0)}{|\Delta\beta_{1}|}\Psi_{10}(t,x,y,z_{c})
\!\!\int_{-\infty}^{\infty} \!\!\!\!\! d\tilde t \, 
\bar\Psi_{20}^{2}(\tilde t,x,y,z_{c}).  
\label{pb17}
\end{eqnarray} 
From Eq. (\ref{pb17}) it follows that inside the collision interval, 
the temporal shape of the pulsed-beam is preserved, while the spatial 
shape is changed by the collision. 
It also follows that one can use the collision to induce 
strong localized changes in the spatial shape of one of the pulsed-beams 
(e.g., pulsed-beam 1) by choosing a second pulsed-beam that 
is spatially localized around the $z$ axis at $z=z_{c}$. 
In the current paper, we use these properties of the fast collision 
to design collision setups that lead to relatively strong changes 
in the spatial shapes of the pulsed-beams even for small values of 
$\epsilon_{3}$ and $1/|\Delta\beta_{1}|$.

The collision-induced change in the shape of pulsed-beam 1 in 
the collision interval $\Delta\Phi_{1}(t,x,y,z_{c})$ is related 
to the collision-induced amplitude shift $\Delta A_{1}^{(c)}$ by: 
\begin{eqnarray}&&
\!\!\!\!\!\!\!\!\!\!\!\!\!\!
\Delta A_{1}^{(c)}=C_{p1}^{-1}
\!\!\int_{-\infty}^{\infty} \!\!\!\!\! dt 
\!\!\int_{-\infty}^{\infty} \!\!\!\!\! dx 
\!\int_{-\infty}^{\infty} \!\!\!\!\! dy
\;\Psi_{10}(t,x,y,z_{c})\Delta\Phi_{1}(t,x,y,z_{c}), 
\label{pb21}
\end{eqnarray}       
where 
\begin{eqnarray}&&
C_{p1}= 
\!\!\int_{-\infty}^{\infty} \!\!\!\!\! dt  
\!\!\int_{-\infty}^{\infty} \!\!\!\!\! dx 
\!\int_{-\infty}^{\infty} \!\!\!\!\! dy
\; \Psi_{10}^{2}(t,x,y,0).  
\label{pb22}
\end{eqnarray}       
Substitution of Eq. (\ref{pb17}) into  Eq. (\ref{pb21}) yields the following   
equation for the collision-induced amplitude shift of pulsed-beam 1 for 
the general initial condition (\ref{pb7}): 
\begin{eqnarray} &&
\!\!\!\!\!\!\!\!\!\!\!\!
\Delta A_{1}^{(c)}=-\frac{2\epsilon_{3}
A_{1}(0) A_{2}^{2}(0)}{C_{p1}|\Delta\beta_{1}|}
%\nonumber \\&&
%\times
\!\int_{-\infty}^{\infty} \!\!\!\!\! dt  
\!\int_{-\infty}^{\infty} \!\!\!\!\! dx 
\!\int_{-\infty}^{\infty} \!\!\!\!\! dy
\;\Psi_{10}^{2}(t,x,y,z_{c})
\!\int_{-\infty}^{\infty} \!\!\!\!\! d\tilde t \;\bar\Psi_{20}^{2}(\tilde t,x,y,z_{c}). 
\label{pb23}
\end{eqnarray}  
Note that Eq.  (\ref{pb23}) for $\Delta A_{1}^{(c)}$ contains integrals 
with respect to $x$ and $y$, while Eq. (\ref{pb17}) for 
$\Delta\Phi_{1}(t,x,y,z_{c})$ does not. As a result, 
it is possible to find collision setups, which lead to strong localized 
changes in the spatial shape of one of the pulsed-beams (e.g., pulsed-beam 1), 
and for which the collision-induced amplitude shift is relatively small.

\subsubsection{Calculation of $\phi_{1}(t,x,y,z)$ in the post-collision interval}  

In the post-collision interval, $z > z_{c} + \Delta z_{c}/2$, 
the pulsed-beams are no longer overlapping. Consequently, the   
nonlinear interaction terms $-2i\epsilon_{3}|\psi_{2}|^2\psi_{1}$ and 
$-2i\epsilon_{3}|\psi_{1}|^2\psi_{2}$ are negligible in this interval.  
Therefore, in the leading order of the perturbation theory, 
the equation for $\phi_{1}$ in the post-collision interval is 
the unperturbed linear propagation equation 
\begin{eqnarray}&&
i\partial_z\phi_{1} 
- \mbox{sgn}(\tilde\beta_{21}) \partial^{2}_{t} \phi_{1} 
+ d_{21}\partial_{x}^{2}\phi_{1}
+ d_{21}\partial_{y}^{2}\phi_{1}=0. 
\label{pb25}
\end{eqnarray}       
In a fast collision $|\Delta\beta_{1}| \gg 1$, and as a result, 
$\Delta\Phi_{1}(t,x,y,z_{c})\simeq \Phi_{1}(t,x,y,z_{c}^{+})$, 
where $\Phi_{1}(t,x,y,z_{c}^{+})$ is the limit 
from the right of $\Phi_{1}(t,x,y,z)$ at $z=z_{c}$.
It follows that the initial condition for Eq. (\ref{pb25}) is:
\begin{eqnarray}&&
\phi_{1}(t,x,y,z_{c}^{+}) = \Phi_{1}(t,x,y,z_{c}^{+})\exp[i\chi_{10}(t,x,y,z_{c})],  
\label{pb26}
\end{eqnarray}        
where $\Phi_{1}(t,x,y,z_{c}^{+})$ is given by Eq. (\ref{pb17}).     
The solution of Eq. (\ref{pb25}) with the initial condition (\ref{pb26}) is  
\begin{eqnarray}&&
\!\!\!\!\!\!\!\!\!\!\!
\phi_{1}(t,x,y,z) \!=\! {\cal F}^{-1} \! \left(\hat\phi_{1}(\omega,k_{1},k_{2},z_{c}^{+})
\exp\{i [\mbox{sgn}(\tilde\beta_{21})\omega^{2} \! 
-d_{21}k_{1}^{2} \! -d_{21}k_{2}^{2}](z-z_{c})\}\right) \!,  
%\nonumber \\&&
\label{pb27}
\end{eqnarray}         
where $\hat\phi_{1}(\omega,k_{1},k_{2},z_{c}^{+})=
{\cal F}\left(\phi_{1}(t,x,y,z_{c}^{+}) \right)$, 
and ${\cal F}$ and ${\cal F}^{-1}$ denote the Fourier transform and the 
inverse Fourier transform with respect to $t$, $x$, and $y$.    
We can write $\phi_{1}(t,x,y,z)$ in the form 
$\phi_{1}(t,x,y,z)=|\phi_{1}(t,x,y,z)|\exp[i \chi_{1}^{(tot)}(t,x,y,z)]$,  
where $\chi_{1}^{(tot)}(t,x,y,z)$ is a real-valued phase factor. 
In general, $\chi_{1}^{(tot)}(t,x,y,z) \ne \chi_{10}(t,x,y,z)$ 
inside the post-collision interval. We therefore define the 
difference between the phase factors of $\psi_{10}$ 
and $\phi_{1}$ by: 
\begin{equation}
\Delta\chi_{1}^{(tot)}(t,x,y,z)= 
\chi_{10}(t,x,y,z) - \chi_{1}^{(tot)}(t,x,y,z). 
\label{pb28}
\end{equation}

\subsection{The predictions of the perturbation approach 
for a separable initial condition}
\label{th_separable}

Let us briefly describe the perturbation theory predictions  
for the collision-induced effects in the important case, 
where the initial condition is given by Eq. (\ref{pb8}),  
which is separable for both pulsed-beams. This case is of special 
interest for two reasons. First, this initial condition corresponds 
to the output electric field from many types of lasers \cite{Siegman86,Kogelnik66}.   
Second, in this case, it is possible to simplify the expressions for the collision-induced 
changes in the shape and amplitude of the pulsed-beams even further, and by this, 
obtain deeper insight into the collision dynamics. For brevity, we present here only 
the end results of the calculations for $\Delta\Phi_{1}(t,x,y,z_{c})$, 
$\Delta A_{1}^{(c)}$, and $\phi_{1}(t,x,y,z)$. We refer the interested 
reader to Ref. \cite{PHN2022} for the details of the derivations that 
were carried out for time-independent beams.

We first note that the solutions $\psi_{j0}$ of the unperturbed linear 
propagation equations (\ref{pb12}) and (\ref{pb13}) with 
the separable initial condition (\ref{pb8}) and with unit amplitude 
can be written as:   
\begin{eqnarray}&&
\psi_{j0}(t,x,y,z)=g_{j}^{(t)}(t,z)g_{j}^{(x)}(x,z)g_{j}^{(y)}(y,z)
\exp\left(i\alpha_{j0} \right),
\label{pb32}
\end{eqnarray}     
where 
\begin{eqnarray}&&
\!\!\!\!\!\!\!\!\!\!\!\!
g_{j}^{(t)}(t,z)=G_{j}^{(t)}(t,z)\exp\left[i\chi_{j0}^{(t)}(t,z)\right], \;\;
g_{j}^{(x)}(x,z)=G_{j}^{(x)}(x,z)\exp\left[i\chi_{j0}^{(x)}(x,z)\right], 
\nonumber \\&&
g_{j}^{(y)}(y,z)=G_{j}^{(y)}(y,z)\exp\left[i\chi_{j0}^{(y)}(y,z)\right],
\label{pb32_add1}
\end{eqnarray}    
and $G_{j}^{(t)}(t,z)$, $G_{j}^{(x)}(x,z)$, $G_{j}^{(y)}(y,z)$, 
$\chi_{j0}^{(t)}(t,z)$, $\chi_{j0}^{(x)}(x,z)$, 
and $\chi_{j0}^{(y)}(y,z)$ are real-valued.  
These solutions can also be expressed in the form 
\begin{eqnarray}&&
\psi_{j0}(t,x,y,z)=\Psi_{j0}(t,x,y,z)
\exp\left[i\chi_{j0}(t,x,y,z) \right],
\label{pb32_add2}
\end{eqnarray}     
where $\Psi_{j0}(t,x,y,z)=G_{j}^{(t)}(t,z)G_{j}^{(x)}(x,z)G_{j}^{(y)}(y,z)$, 
and 
\begin{eqnarray}&&
\chi_{j0}(t,x,y,z)=\chi_{j0}^{(t)}(t,z) + \chi_{j0}^{(x)}(x,z) 
+ \chi_{j0}^{(y)}(y,z) + \alpha_{j0}  
\label{pb32_add3}
\end{eqnarray}     
is the overall phase factor.

For a separable initial condition, one can use the conservation 
of the total energy for the unperturbed linear propagation equation 
to further simplify the expressions for the collision-induced 
changes in the pulsed-beam's amplitude and shape. Using this 
conservation law together with Eq. (\ref{pb8}), we find   
\begin{eqnarray}&&
\!\!\!\!\!\!\!\!\!\!\!\!\!\!\!\!\!\!\!\!
\int_{-\infty}^{\infty} \!\!\!\! dt \, G_{j}^{(t)2}(t,z)
=\int_{-\infty}^{\infty} \!\!\!\! dt \, G_{j}^{(t)2}(t,0)
=W_{j0}^{(t)}\int_{-\infty}^{\infty} \!\!\!\! ds \, h_{j}^{(t)2}(s)
=W_{j0}^{(t)}  c_{pj}^{(t)},    
\label{pb33}
\end{eqnarray} 
\begin{eqnarray}&&
\!\!\!\!\!\!\!\!\!\!\!\!\!\!\!\!\!\!\!\!
\int_{-\infty}^{\infty} \!\!\!\! dx \, G_{j}^{(x)2}(x,z)
=\int_{-\infty}^{\infty} \!\!\!\! dx \, G_{j}^{(x)2}(x,0)
=W_{j0}^{(x)}\int_{-\infty}^{\infty} \!\!\!\! ds \, h_{j}^{(x)2}(s)
=W_{j0}^{(x)}  c_{pj}^{(x)},  
\label{pb33B}
\end{eqnarray}         
and 
\begin{eqnarray}&&
\!\!\!\! \!\!\!\!\!\!\!\!\!\!\!\!\!\!\!\!
\int_{-\infty}^{\infty} \!\!\!\! dy \, G_{j}^{(y)2}(y,z)
=\int_{-\infty}^{\infty} \!\!\!\! dy \, G_{j}^{(y)2}(y,0)
=W_{j0}^{(y)}\int_{-\infty}^{\infty} \!\!\!\! ds \, h_{j}^{(y)2}(s)
=W_{j0}^{(y)}  c_{pj}^{(y)}.   
\label{pb33C}
\end{eqnarray}               
Equations (\ref{pb33})-(\ref{pb33C}) define the constants 
$c_{pj}^{(t)}$, $c_{pj}^{(x)}$, and $c_{pj}^{(y)}$, which 
appear in the expressions for $\Delta A_{1}^{(c)}$, 
$\Delta\Phi_{1}$, and $\phi_{1}$.

\subsubsection{Collision-induced effects in the collision interval}  

In the case of a separable initial condition, one can simplify Eq. (\ref{pb17}) 
for the collision-induced change in the shape of pulsed-beam 1 
in the collision interval to the following form: 
\begin{eqnarray} &&
\!\!\!\!\!\!\!
\Delta\Phi_{1}(t,x,y,z_{c})\!=\!-\frac{2\epsilon_{3}
A_{1}(0) A_{2}^{2}(0)}{|\Delta\beta_{1}|}
c_{p2}^{(t)} W_{20}^{(t)}
\nonumber \\&&
\times G_{1}^{(t)}(t,z_{c}) 
G_{1}^{(x)}(x,z_{c})G_{1}^{(y)}(y,z_{c})
G_{2}^{(x)2}(x,z_{c})G_{2}^{(y)2}(y,z_{c}).  
\label{pb31}
\end{eqnarray}    
From Eq. (\ref{pb31}) it follows that when the initial condition is separable, 
the shape of the pulsed-beam does not change at all due to the collision 
(see also subsection \ref{lp_separable_PC}). Moreover, as will be demonstrated 
in section \ref{setups}, the simple form of Eq. (\ref{pb31}) can be exploited for designing 
collision setups that lead to localized and nonlocalized strong intensity reduction effects.

The collision-induced amplitude shift for pulsed-beam 1  
in the case of a separable initial condition is given by: 
\begin{eqnarray} &&
\!\!\!\!
\Delta A_{1}^{(c)}=-\frac{2\epsilon_{3}
A_{1}(0) A_{2}^{2}(0)}{|\Delta\beta_{1}|}
\frac{c_{p2}^{(t)}W_{20}^{(t)}} 
{c_{p1}^{(x)}W_{10}^{(x)}c_{p1}^{(y)}W_{10}^{(y)}}
\nonumber \\&&
\times
\!\int_{-\infty}^{\infty} \!\!\!\!\! dx \, 
G_{1}^{(x)2}(x,z_{c})G_{2}^{(x)2}(x,z_{c})
\!\int_{-\infty}^{\infty} \!\!\!\!\! dy \, 
G_{1}^{(y)2}(y,z_{c})G_{2}^{(y)2}(y,z_{c}). 
\label{pb34}
\end{eqnarray}       
We observe that Eq. (\ref{pb34}) has the form 
\begin{eqnarray} &&
\!\!\!\!\!\!\!\!\!\!\!\!\!\!\!\!\!\!\!\!
\Delta A_{1}^{(c)}=-(\mbox{overall factor}) \times 
(\mbox{temporal factor}) \times (\mbox{spatial factor}), 
\label{pb35}
\end{eqnarray}  
where the overall factor is $2\epsilon_{3}A_{1}(0) A_{2}^{2}(0)/|\Delta\beta_{1}|$, 
and the temporal factor is $c_{p2}^{(t)}W_{20}^{(t)}$. 
Furthermore, the temporal factor is universal in the sense that it 
does not depend on the exact details of the initial pulsed-beam shapes 
and on the collision distance $z_{c}$.

\subsubsection{Calculation of $\phi_{1}(t,x,y,z)$ in the post-collision interval}
\label{lp_separable_PC}

We now turn to describe the calculation of $\phi_{1}(t,x,y,z)$ in the 
post-collision interval for a separable initial condition. The dynamics 
of $\phi_{1}(t,x,y,z)$ is still described by Eq. (\ref{pb25}). 
The initial condition for Eq. (\ref{pb25}) can now be written as 
\begin{eqnarray} &&
\phi_{1}(t,x,y,z_{c}^{+})=
-\tilde a_{1} g_{1}^{(t)}(t,z_{c})
g_{12}^{(x)}(x,z_{c})g_{12}^{(y)}(y,z_{c})\exp(i\alpha_{10}),
\label{pb40}
\end{eqnarray}        
where 
\begin{eqnarray} &&
\tilde a_{1}=
2\epsilon_{3}A_{1}(0) A_{2}^{2}(0) 
c_{p2}^{(t)} W_{20}^{(t)}/|\Delta\beta_{1}|, 
\label{pb41}
\end{eqnarray}        
and 
\begin{eqnarray} &&
\!\!\!\!\!\!\!\!\!\!\!\!\!\!\!\!\!\!\!\!\!
g_{12}^{(x)}(x,z_{c})= 
g_{1}^{(x)}(x,z_{c})G_{2}^{(x)2}(x,z_{c}), 
\;\;\;\;
g_{12}^{(y)}(y,z_{c})= 
g_{1}^{(y)}(y,z_{c})G_{2}^{(y)2}(y,z_{c}).
\label{pb42}
\end{eqnarray}        
It is then possible to show that the expression for $\phi_{1}(t,x,y,z)$ 
in the post-collision interval is 
\begin{eqnarray}&&
\phi_{1}(t,x,y,z)=
-\tilde a_{1} g_{1}^{(t)}(t,z)
{\cal F}^{-1}\left(\hat g_{12}^{(x)}(k_{1},z_{c})
\exp[-i d_{21} k_{1}^{2}(z-z_{c})]\right)
\nonumber \\&&
\times  
{\cal F}^{-1}\left(\hat g_{12}^{(y)}(k_{2},z_{c})
\exp[-i d_{21} k_{2}^{2}(z-z_{c})]\right)
\exp(i\alpha_{10}), 
\label{pb43}
\end{eqnarray}       
where $\hat g_{12}^{(x)}(k_{1},z_{c})$ and $\hat g_{12}^{(y)}(k_{2},z_{c})$ 
are the Fourier transforms of $g_{12}^{(x)}(x,z)$ and $g_{12}^{(y)}(y,z)$ 
with respect to $x$ and $y$, respectively.      
We observe that when the initial condition is separable for both pulsed-beams  
the $t$ dependences of $\phi_{1}(t,x,y,z)$ and $\psi_{10}(t,x,y,z)$ 
are identical for $z>z_{c}$. It follows that in this case,  
the temporal shape of the pulsed-beam is not changed by 
the collision at all. In contrast, the spatial shape of the pulsed-beam 
is changed by the collision, and this change is proportional to the 
product of the two inverse Fourier transforms on the right hand side 
of Eq. (\ref{pb43}).

The real-valued phase factor $\chi_{1}^{(tot)}(t,x,y,z)$ associated with 
$\phi_{1}(t,x,y,z)$ can be written as 
\begin{equation}
\chi_{1}^{(tot)}(t,x,y,z)= 
\chi_{10}^{(t)}(t,z) + \chi_{1}^{(x)}(x,z) + \chi_{1}^{(y)}(y,z) + \alpha_{10} + \pi,  
\label{pb44}
\end{equation}   
where $\chi_{1}^{(x)}(x,z)$ and $ \chi_{1}^{(y)}(y,z)$ are the phase factors 
associated with \linebreak ${\cal F}^{-1}\left(\hat g_{12}^{(x)}(k_{1},z_{c})
\exp[-i d_{21} k_{1}^{2}(z-z_{c})]\right)$ and 
${\cal F}^{-1}\left(\hat g_{12}^{(y)}(k_{2},z_{c})
\exp[-i d_{21} k_{2}^{2}(z-z_{c})]\right)$, respectively. 
Using Eqs. (\ref{pb32_add3}) and (\ref{pb44}) we obtain that in the case 
of a separable initial condition, the phase factor difference 
$\Delta\chi_{1}^{(tot)}$ between $\psi_{10}$ and $\phi_{1}$ 
is given by: 
\begin{equation}
\Delta\chi_{1}^{(tot)}(t,x,y,z)= 
\chi_{10}^{(x)}(x,z) + \chi_{10}^{(y)}(y,z) 
- \chi_{1}^{(x)}(x,z) - \chi_{1}^{(y)}(y,z) - \pi.   
\label{pb44_add1}
\end{equation}

\section{Collision setups that lead to strong intensity reduction effects}
\label{setups}

In this section, we describe the guiding principles for design of collision 
setups that lead to strong localized and nonlocalized intensity reduction effects. 
We also present the calculation of the collision-induced change in the 
pulsed-beam's shape and amplitude, and the calculation of the intensity reduction 
factor for these setups.

\subsection{Calculation of $\phi_{1}(t,x,y,z)$ and $\Delta A_{1}^{(c)}$ for Gaussian pulsed-beams}
\label{beam_shape}
We consider a separable initial condition in the form of two 
Gaussian pulsed-beams:  
\begin{eqnarray}&&
\!\!\!\!\!\!\!\!\!\!\!
\psi_{1}(t,x,y,0)=A_{1}(0)\exp \left[-\frac{t^{2}}{2W^{(t)2}_{10}}
-\frac{x^{2}}{2W^{(x)2}_{10}} -\frac{y^{2}}{2W^{(y)2}_{10}} 
+ i\alpha_{10} \right],
\nonumber \\&&
\!\!\!\!\!\!\!\!\!\!\!
\psi_{2}(t,x,y,0)=A_{2}(0)\exp \left[-\frac{(t-t_{20})^{2}}{2W^{(t)2}_{20}}
-\frac{x^{2}}{2W^{(x)2}_{20}} -\frac{y^{2}}{2W^{(y)2}_{20}} 
+ i\alpha_{20} \right]. 
\label{pb45} 
\end{eqnarray}  
We consider this form of the initial pulsed-beam shapes, since it is 
highly accessible for laser-beam propagation experiments \cite{Siegman86,Kogelnik66}. 
In addition, this choice allows us to obtain an explicit formula 
for $\phi_{1}$ in the post-collision interval. We point out   
that Eq. (\ref{pb43}) of our perturbation approach can be used for 
calculating $\phi_{1}$ for general separable pulsed-beam 
shapes of the form (\ref{pb8}). However, the latter calculation  
would require numerical integration of the integrals appearing 
on the right hand side of Eq. (\ref{pb43}).

Since the initial condition is separable, we can calculate $\phi_{1}$  
by employing Eq. (\ref{pb43}). Using Eq. (\ref{pb33}), we find 
$c_{p2}^{(t)}=\pi^{1/2}$ and 
\begin{eqnarray} &&
\tilde a_{1}=
2\pi^{1/2}\epsilon_{3}A_{1}(0) A_{2}^{2}(0)
W_{20}^{(t)}/|\Delta\beta_{1}|.  
\label{pb65}
\end{eqnarray}      
The solution of the unperturbed linear propagation equation 
with the initial condition (\ref{pb45}) yields the following 
expression for $g_{1}^{(t)}(t,z)$: 
\begin{eqnarray}&&
\!\!\!\!\!\!\!\!\!\!\!\!\!\!
g_{1}^{(t)}(t,z)= 
\frac{W_{10}^{(t)}}{(W_{10}^{(t)4} + 4z^{2})^{1/4}}
\exp\left[-\frac{W_{10}^{(t)2} t^{2}}{2(W_{10}^{(t)4} + 4z^{2})}
+i\chi_{10}^{(t)}(t,z)\right], 
\label{pb66}
\end{eqnarray}
where 
\begin{eqnarray} &&
\chi_{10}^{(t)}(t,z)= 
\frac{1}{2}\mbox{sgn}(\tilde\beta_{21})
\arctan\left(\frac{2z}{W_{10}^{(t)2}}\right) 
- \frac{\mbox{sgn}(\tilde\beta_{21}) t^{2}z}{W_{10}^{(t)4} + 4z^{2}}. 
\label{pb67}
\end{eqnarray}   
The two inverse Fourier transforms on the right hand side of 
Eq. (\ref{pb43}) are calculated in Appendix \ref{appendA}. 
The two calculations are similar to one another. Furthermore, 
the results can be expressed in a single formula that has the 
following form: 
\begin{eqnarray}&&
{\cal F}^{-1}\left(\hat g_{12}^{(u)}(k_u,z_c)
\exp[-id_{21} k_u^2(z-z_c)]\right) =
\nonumber\\&&
=\frac{W_{10}^{(u)}W_{20}^{(u)2} 
\exp\left[-q_{1}^{(u)}(z_c)u^2/R_{1}^{(u)4}(z,z_c) 
+ i\chi_1^{(u)}(u,z)\right]}
{(W_{10}^{(u)4} + 4d_{21}^2 z_c^2)^{1/4} 
(W_{20}^{(u)4} + 4d_{22}^2 z_c^2)^{1/2} R_{1}^{(u)}(z,z_c)},
\label{pb68}
\end{eqnarray} 
where the $u$ stands for 1 or 2 in $k_{u}$, and for 
$x$ or $y$, respectively, everywhere else in the formula. 
The quantities $q_{1}^{(u)}(z_c)$, $R_{1}^{(u)}(z,z_c)$, 
and $\chi_1^{(u)}(u,z)$ in Eq. (\ref{pb68}) are given by 
Eqs. (\ref{appA3}), (\ref{appA8}), and (\ref{appA9}) in Appendix \ref{appendA}. 
Substitution of Eqs. (\ref{pb65})-(\ref{pb68}) into Eq. (\ref{pb43}) 
yields the following expression for $\phi_{1}(t,x,y,z)$: 
\begin{eqnarray}&&
\!\!\!\!\!\!\!\!\!\!\!\!\!\!\!\!\!
\phi_{1}(t,x,y,z)= 
\frac{\tilde a_{1}W_{10}^{(t)}W_{10}^{(x)}W_{10}^{(y)}W_{20}^{(x)2}W_{20}^{(y)2}}
{(W_{10}^{(t)4} + 4z^{2})^{1/4}
(W_{10}^{(x)4} + 4d_{21}^2 z_{c}^{2})^{1/4}
(W_{10}^{(y)4} + 4d_{21}^2 z_{c}^{2})^{1/4}}
\nonumber \\&&
\!\!\!\!\!\!\!\!\!\!\!\!\!\!\!\!\!
\times
\left[(W_{20}^{(x)4} + 4d_{22}^{2}z_{c}^{2})^{1/2}
(W_{20}^{(y)4} + 4d_{22}^{2}z_{c}^{2})^{1/2}
R_{1}^{(x)}(z,z_c)R_{1}^{(y)}(z,z_c)
\right]^{-1}
\nonumber \\&& 
\!\!\!\!\!\!\!\!\!\!\!\!\!\!\!\!\!
\times
\exp \! \left[
-\frac{W_{10}^{(t)2} t^{2}}{2(W_{10}^{(t)4} + 4z^{2})}
\!-\! \frac{q_{1}^{(x)}(z_{c}) x^{2}}{R_{1}^{(x)4}(z,z_{c})} 
\!-\! \frac{q_{1}^{(y)}(z_{c}) y^{2}}{R_{1}^{(y)4}(z,z_{c})} 
\!+\! i\chi_{1}^{(tot)}(t,x,y,z)\right] \!\!,  
\label{pb70}     
\end{eqnarray}   
where the total phase factor $\chi_{1}^{(tot)}$ is of the form 
(\ref{pb44}), $\chi_{10}^{(t)}$ is given by Eq. (\ref{pb67}),  
and $\chi_{1}^{(x)}$ and $\chi_{1}^{(y)}$ are given by 
Eq. (\ref{appA9}) in Appendix \ref{appendA}.
We observe that the expression for $\phi_{1}$ is rather complicated 
even for the relatively simple Gaussian initial condition (\ref{pb45}). 
For this reason, Eq. (\ref{pb70}) will not be directly used in the 
design of the collision setups that lead to significant intensity reduction 
effects. Instead, we will use an expression that is based on the much 
simpler form of Eq. (\ref{pb31}) for the collision-induced change in the 
pulsed-beam's shape in the collision-interval.

The equation for the collision-induced amplitude shift $\Delta A_1^{(c)}$  
is obtained by employing Eq. (\ref{pb34}). Using the expressions for 
the solutions of the unperturbed linear propagation equation with the 
initial condition (\ref{pb45}) in Eq. (\ref{pb34}), we find: 
\begin{eqnarray}&&
\Delta A_1^{(c)} = \frac{-2\pi^{1/2}\epsilon_3 A_1(0)A_2^2(0)}{|\Delta\beta_{1}|}
W_{10}^{(x)}W_{10}^{(y)} W_{20}^{(t)}W_{20}^{(x)2}W_{20}^{(y)2}
\nonumber\\&&
\times
\left[W_{10}^{(x)2}(W_{20}^{(x)4} + 4d_{22}^{2}z_{c}^{2})
+W_{20}^{(x)2}(W_{10}^{(x)4} + 4d_{21}^{2}z_{c}^{2})
\right]^{-1/2}  
\nonumber\\&&
\times
\left[W_{10}^{(y)2}(W_{20}^{(y)4} + 4d_{22}^{2}z_{c}^{2})
+W_{20}^{(y)2}(W_{10}^{(y)4} + 4d_{21}^{2}z_{c}^{2})
\right]^{-1/2}.  
 \!\!\!\!\!\!\!\!\!\!\!\!\!\!
\label{pb71}
\end{eqnarray}

\subsection{The fractional intensity reduction factor}    

The main physical quantity, which is used for estimating the strength 
of the collisional effects, is the fractional intensity reduction factor. 
We therefore discuss here the definition and the basic properties of 
this quantity. The fractional intensity reduction factor for pulsed-beam 1  
$\Delta I_1^{(r)}$ is defined by: 
\begin{eqnarray}&& 
\!\!\!\!\!\!\!\!\!\!\!\!\!\!\!\!\!\!\!\!\!
\Delta I_1^{(r)}(t,x,y,z)=
\frac{\tilde I_{1}(t,x,y,z)-I_{1}(t,x,y,z)}{\tilde I_{1}(t,x,y,z)}=
1-\frac{I_{1}(t,x,y,z)}{A_{1}^{2}(0) \Psi_{10}^{2}(t,x,y,z)},
\label{pb46}
\end{eqnarray} 
where $I_{1}(t,x,y,z)=|\psi_{1}(t,x,y,z)|^{2}$ is the intensity of pulsed-beam 1 
in the presence of cubic loss, and $\tilde I_{1}(t,x,y,z)=A_{1}^{2}(0) 
\Psi_{10}^{2}(t,x,y,z)$ is the intensity of pulsed-beam 1 in the absence 
of cubic loss. It follows that $\Delta I_1^{(r)}$ measures 
the ratio between the intensity decrease of pulsed-beam 1, which is caused  
by the effects of cubic loss on the collision, and the intensity 
of the pulsed-beam for unperturbed single-beam propagation.

The approximate prediction of the perturbation theory for the fractional 
intensity reduction factor, which we denote by $\Delta I_1^{(r)(1)}$,  
is obtained by substituting Eq. (\ref{pb11}) into Eq. (\ref{pb46}) and by 
expanding the result up to order $\epsilon_{3}/|\Delta\beta_{1}|$. This 
calculation yields: 
\begin{eqnarray}&&
\!\!\!\!\!\!\!\!\!\!\!\!\!\!\!\!\!\!\!\!\!
\Delta I_{1}^{(r)(1)}(t,x,y,z) \!=\! 
-\frac{\psi_{10}(t,x,y,z)\phi_{1}^{*}(t,x,y,z) \!+\! 
\psi_{10}^{*}(t,x,y,z)\phi_{1}(t,x,y,z)}
{A_{1}(0) \Psi_{10}^{2}(t,x,y,z)}.
\label{pb47}
\end{eqnarray}      
Using the relations $\psi_{10}=\Psi_{10}\exp[i\chi_{10}]$  
and $\phi_{1}=|\phi_{1}|\exp[i\chi_{1}^{(tot)}]$, we obtain: 
\begin{eqnarray}&&
\Delta I_{1}^{(r)(1)}(t,x,y,z)= 
-\frac{2|\phi_{1}(t,x,y,z)|\cos\left[\Delta \chi_{1}^{(tot)}(t,x,y,z)\right]}
{A_{1}(0) \Psi_{10}(t,x,y,z)}.
\label{pb48}
\end{eqnarray}
For a separable initial pulsed-beam input, $\Delta \chi_{1}^{(tot)}$ 
is given by Eq. (\ref{pb44_add1}). In addition, the $t$ dependences of $\Psi_{10}$ 
and $|\phi_{1}|$ are identical. Therefore, in this case, the dependence 
on $t$ cancels out on the right hand side of Eq. (\ref{pb48}), 
and $\Delta I_1^{(r)(1)}$ becomes independent of $t$. It follows that  
for a separable initial pulsed-beam input, the expression for 
$\Delta I_1^{(r)(1)}$ is: 
\begin{eqnarray}&&
\!\!\!\!\!\!\!\!\!\!\!\!\!\!\!\!\!\!\!\!\!
\Delta I_{1}^{(r)}(x,y,z)= 
-\frac{2|\phi_{1}(t,x,y,z)|}
{A_{1}(0) \Psi_{10}(t,x,y,z)}
\nonumber \\&&
\times  
\cos\left[\chi_{10}^{(x)}(x,z) + \chi_{10}^{(y)}(y,z) 
-\chi_{1}^{(x)}(x,z) - \chi_{1}^{(y)}(y,z)-\pi \right].  
\label{pb49}
\end{eqnarray}

The form of the approximate expression for the intensity reduction 
factor at $z=z_{c}^{+}$ is of particular interest, since it is used for the design of  
collision setups that lead to strong intensity reduction effects. 
We therefore turn to obtain the expression for $\Delta I_{1}^{(r)(1)}(x,y,z_{c}^{+})$. 
From Eq. (\ref{pb26}) it follows that $|\phi_{1}(t,x,y,z_{c}^{+})|= 
\Phi_{1}(t,x,y,z_{c}^{+})=\Delta\Phi_{1}(t,x,y,z_{c})$
and $\Delta \chi_{1}^{(tot)}(x,y,z_{c}^{+})=0$. 
Therefore, the expression for $\Delta I_{1}^{(r)(1)}(x,y,z_{c}^{+})$ is 
\begin{eqnarray}&&
\!\!\!\!\!\!\!\!\!\!\!\!\!\!\!\!
\Delta I_{1}^{(r)(1)}(x,y,z_{c}^{+}) = 
-\frac{2\Delta\Phi_{1}(t,x,y,z_{c})}
{A_{1}(0) \Psi_{10}(t,x,y,z_{c})}.
\label{pb50}
\end{eqnarray}      
When the initial pulsed-beam input is separable, we can express 
$\Delta\Phi_{1}$ by Eq. (\ref{pb31}) and obtain: 
\begin{eqnarray} &&
\!\!\!\!\!\!\!
\Delta I_{1}^{(r)(1)}(x,y,z_{c}^{+}) = 
\frac{4\epsilon_{3}A_{2}^{2}(0)}{|\Delta\beta_{1}|}
c_{p2}^{(t)} W_{20}^{(t)} G_{2}^{(x)2}(x,z_{c}) G_{2}^{(y)2}(y,z_{c}). 
\label{pb51}
\end{eqnarray} 
We take advantage of the simple form of Eq. (\ref{pb51}), and use it 
to estimate the physical parameter values required for realizing strong 
collision-induced intensity reduction effects.

\subsection{Design of the collision setups}

The design of collision setups that lead to strong localized 
and nonlocalized intensity reduction effects is based on 
the simple form of Eq. (\ref{pb51}). We employ this equation 
for the Gaussian initial condition (\ref{pb45}) and obtain 
\begin{eqnarray} &&
\!\!\!\!\!\!\!
\Delta I_{1}^{(r)(1)}(x,y,z_{c}^{+}) = 
\frac{4\pi^{1/2}\epsilon_{3}A_{2}^{2}(0) W_{20}^{(t)} W_{20}^{(x)} W_{20}^{(y)}}
{|\Delta\beta_{1}| W_{2}^{(x)}(z_{c}) W_{2}^{(y)}(z_{c})} 
%\nonumber \\&&
%\times 
\exp \left[ -\frac{x^{2}}{W_{2}^{(x)2}(z_{c})} 
-\frac{y^{2}}{W_{2}^{(y)2}(z_{c})} \right],  
\label{pb52}
\end{eqnarray}   
where the $z$-dependent pulsed-beam widths in the $x$ and $y$ directions 
are given by 
\begin{equation}
W_{j}^{(x)}(z)=\left(W_{j0}^{(x)2} + \frac{4d_{2j}^{2}z^{2}}{W_{j0}^{(x)2}}\right)^{1/2}, \; 
W_{j}^{(y)}(z)=\left(W_{j0}^{(y)2} + \frac{4d_{2j}^{2}z^{2}}{W_{j0}^{(y)2}}\right)^{1/2}.  
\label{pb53}
\end{equation}
Further simplification is obtained by calculating $\Delta I_{1}^{(r)}$ 
on the $z$ axis. We obtain:   
\begin{eqnarray} &&
\!\!\!\!\!\!\!
\Delta I_{1}^{(r)(1)}(0,0,z_{c}^{+}) = 
\frac{4\pi^{1/2}\epsilon_{3}A_{2}^{2}(0) W_{20}^{(t)} W_{20}^{(x)} W_{20}^{(y)}}
{|\Delta\beta_{1}| W_{2}^{(x)}(z_{c}) W_{2}^{(y)}(z_{c})}.    
\label{pb54}
\end{eqnarray}

We now describe the main guiding principles that are used in  
the design of the collision setups that lead to strong intensity 
reduction effects. 
\begin{enumerate}
\item We consider setups that lead to significant collision-induced intensity 
change but that are not completely outside of the region of validity 
of the perturbative calculation. We therefore require: 
\begin{equation}
 0.1 < \Delta I_{1}^{(r)(1)}(0,0,z_{c}^{+}) < 1.0.  
\label{pb55}
\end{equation}
  
\item (a) For localized collision setups, we look for a generic case,  
in which the diffraction-induced broadening of pulsed-beam 2 at $z=z_{c}$ 
is not very large on one hand, and is not negligible on the other hand. 
We therefore require: 
\begin{equation}
W_{20}^{(x)2} \gtrsim   2d_{22}z_{c} , \;\;\;
W_{20}^{(y)2} \gtrsim   2d_{22}z_{c} . 
\label{pb56}
\end{equation}
We then determine the value of $d_{22}$ by using 
\begin{equation}
d_{22} \approx W_{20}^{(x)2}/(2z_{c}) , \;\;\;  
d_{22} \approx W_{20}^{(y)2}/(2z_{c}) .   
\label{pb57}
\end{equation} 
We emphasize, however, that this condition is not very limiting, 
as somewhat smaller or larger values of $d_{22}$ can be used. 
 
(b) For nonlocalized collision setups, we look for a generic case,  
in which the diffraction-induced broadening of both pulsed-beams at $z=z_{c}$ 
is not very large and not negligible.  
 
\item Spatially localized reduction in the intensity of pulsed-beam 1 
can be realized by requiring that 
\begin{equation}
W_{2}^{(x)}(z_{c}) \ll  W_{1}^{(x)}(z_{c}) , \;\;\; 
W_{2}^{(y)}(z_{c}) \ll  W_{1}^{(y)}(z_{c}) .
\label{pb58}
\end{equation}
Using Eq. (\ref{pb53}), we find that condition (\ref{pb58}) 
can be satisfied by choosing 
\begin{equation}
W_{20}^{(x)} \ll  W_{10}^{(x)} , \;\;\; 
W_{20}^{(y)} \ll  W_{10}^{(y)} .
\label{pb59}
\end{equation} 
In a similar manner, nonlocalized intensity reduction 
for pulsed-beam 1 can be realized by taking 
\begin{equation}
W_{20}^{(x)} \approx  W_{10}^{(x)} , \;\;\; 
W_{20}^{(y)} \approx  W_{10}^{(y)} .
\label{pb60}
\end{equation}   
    
\end{enumerate}

\section{Numerical simulations}
\label{simu}

The predictions of the previous sections are based on a number 
of simplifying approximations, and the validity conditions for 
these approximations are not known. For example, it is unclear 
if the approximation of $\Psi_{20}(t,x,y,z)$ by 
$\bar\Psi_{20}(\tilde t,x,y,z_{c})$, which was used in the derivation 
of Eq. (\ref{pb17}) from Eq. (\ref{pb16}), is valid under the 
conditions specified in the end of subsection \ref{th_intro}.   
Moreover, in the current paper, we develop and study collision setups 
that lead to relatively strong intensity reduction effects. 
For these setups, the validity of the entire 
perturbative calculation might break down. 
For this reason, it is important to check the perturbation theory 
predictions by numerical simulations with Eq. (\ref{pb5}).       
In the current paper, we take on this important investigation. 
More specifically, we carry out numerical simulations with 
Eq. (\ref{pb5}) for two collision setups, one that leads to 
strong localized intensity reduction effects (section \ref{localized}), 
and another that leads to strong nonlocalized intensity reduction 
effects (section \ref{nonlocalized}). In both cases, we numerically 
solve Eq. (\ref{pb5}) with the initial condition (\ref{pb45})   
by the split-step method with periodic boundary 
conditions \cite{Agrawal2019,Yang2010}. Furthermore, we compare 
the simulations results with the approximate predictions 
of the perturbation theory.

\subsection{A collision setup with localized effects}
\label{localized}

We start by considering collision setup 1, which leads to strong 
spatially localized intensity reduction effects. The main conditions that 
guide the choice of the physical parameter values for this setup 
are specified by Eqs. (\ref{pb54}), (\ref{pb55}), (\ref{pb57}), and (\ref{pb59}). 
We use these conditions in the following manner. 

\begin{enumerate}

\item Spatially localized intensity reduction for pulsed-beam 1 
is realized by choosing $W_{10}^{(x)}=W_{10}^{(y)}=5$ and  
$W_{20}^{(x)}=W_{20}^{(y)}=0.5$, in accordance with condition (\ref{pb59}). 

\item We choose $t_{10}=0$ and $t_{20}=-20$, such that the pulsed-beams 
are well-separated at $z=0$. In addition, $\Delta\beta_{1}=10$, 
and as a result, the collision distance is $z_{c}=2$. 
Using condition (\ref{pb57}), we find that $d_{22} \approx 0.0625$, 
so we choose $d_{22}=0.06$. 
   
\item From the relations $d_{2j}= \tau_{0}^{2}\lambda_{j}/
(2\pi |\tilde \beta_{21}|x^{\prime 2}_{0})$ in Eq. (\ref{pb6}), 
it follows that $d_{21}=\lambda_{1}d_{22}/\lambda_{2}$. 
For a realistic choice of the wavelengths (see below), 
$\lambda_{1}$ is close to $\lambda_{2}$, and as a result, 
$d_{21} \approx d_{22}$. We therefore use $d_{21}=0.06$. 

\item Using Eq. (\ref{pb53}) with the parameter values specified 
in the previous items, we find $W_{2}^{(x)}(z_{c})=W_{2}^{(y)}(z_{c}) \simeq 0.693$, 
and $W_{1}^{(x)}(z_{c})=W_{1}^{(y)}(z_{c}) \simeq 5.000$. 
Therefore, condition (\ref{pb58}) is satisfied for the 
chosen parameter values.             
 
\item Strong intensity reduction effects are realized by choosing 
$\epsilon_{3}=0.12$, $A_{2}(0)=2$, and $W_{20}^{(t)}=5$, 
in accordance with Eqs. (\ref{pb54}) and (\ref{pb55}).   
With this choice we obtain $\Delta I_{1}^{(r)(1)}(0,0,z_{c}^{+}) \simeq 0.886$.    

\item The values of the other physical parameters are taken as 
$\beta_{22}=1$, $A_{1}(0)=1$, and $W_{10}^{(t)}=1$. 
Additionally, the final propagation distance is chosen as $z_{f}=2z_{c}=4$, 
such that the pulsed-beams are also well-separated at $z=z_{f}$.  
  
\end{enumerate}

Let us provide an example for the values of the dimensional 
physical parameters that correspond to the values of the dimensionless 
parameters specified in items 1-6. For this purpose, we consider 
bulk fused silica as an example for the optical medium. 
We use the information in Refs. \cite{Malitson65,Tan98} 
and choose $x^{\prime}_{0}=2$ cm, $\lambda_{1}=2.791$ $\mu$m, 
$\lambda_{2}=2.795$ $\mu$m, $\tilde \beta_{21}=-394.23$ 
$\mbox{ps}^{2}\mbox{km}^{-1}$, $\tilde \beta_{22}=-396.59$ 
$\mbox{ps}^{2}\mbox{km}^{-1}$, $\tilde d_{21}=2.221\times 10^{-7}$m, 
$\tilde d_{22}=2.224\times 10^{-7}$m, $\tau_{0}=4.615$ ps, 
$P_{0}=0.05$ W, $\rho_{3}=22.210$ $\mbox{W}^{-1}\mbox{km}^{-1}$,        
and $z''_{c}=216.0$ m \cite{parameters1}.

% fig 1 - pulsed-beams shapes - setup 1  
\begin{figure}[ptb]
\begin{center}
\epsfxsize=10cm  \epsffile{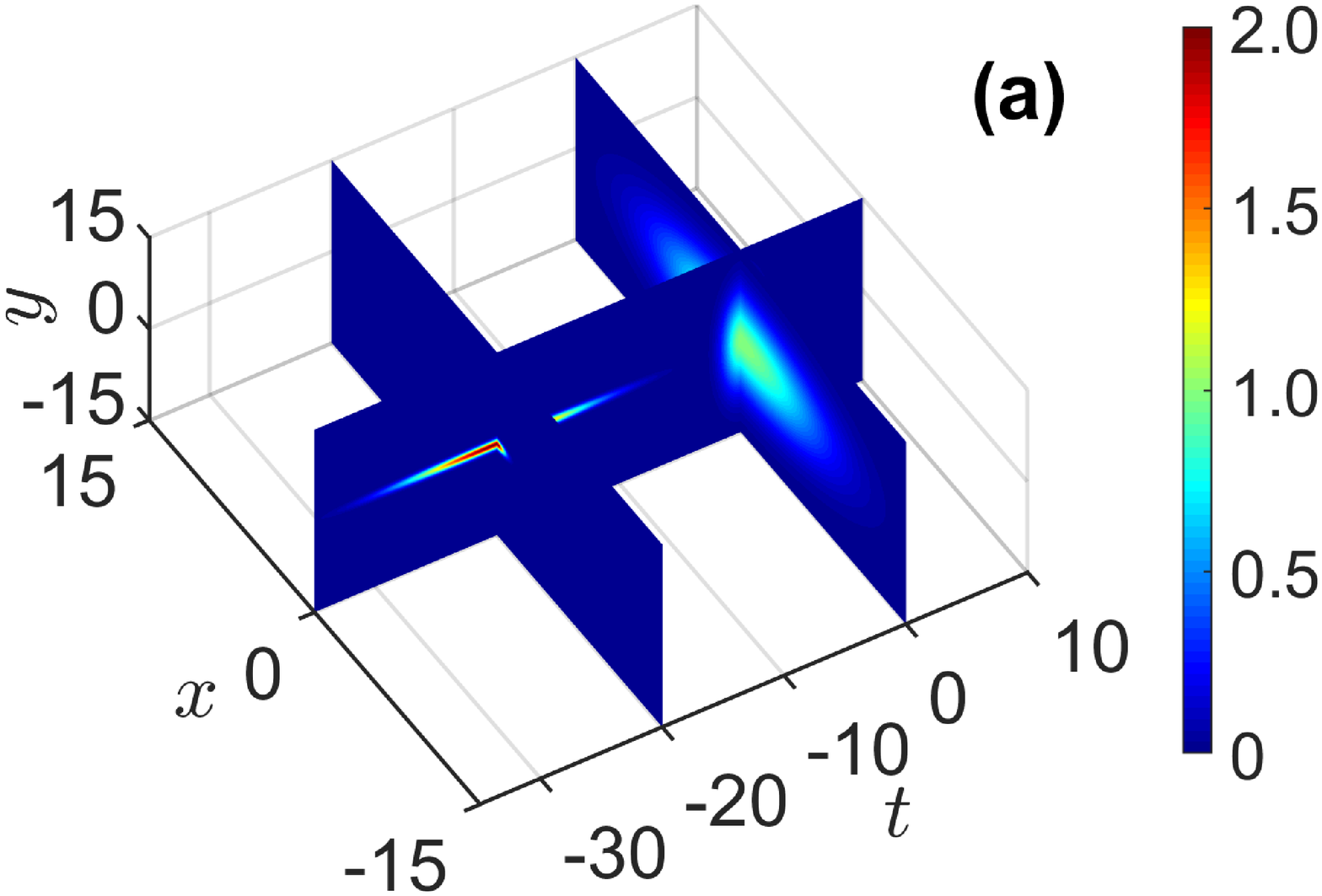}\\
\epsfxsize=10cm  \epsffile{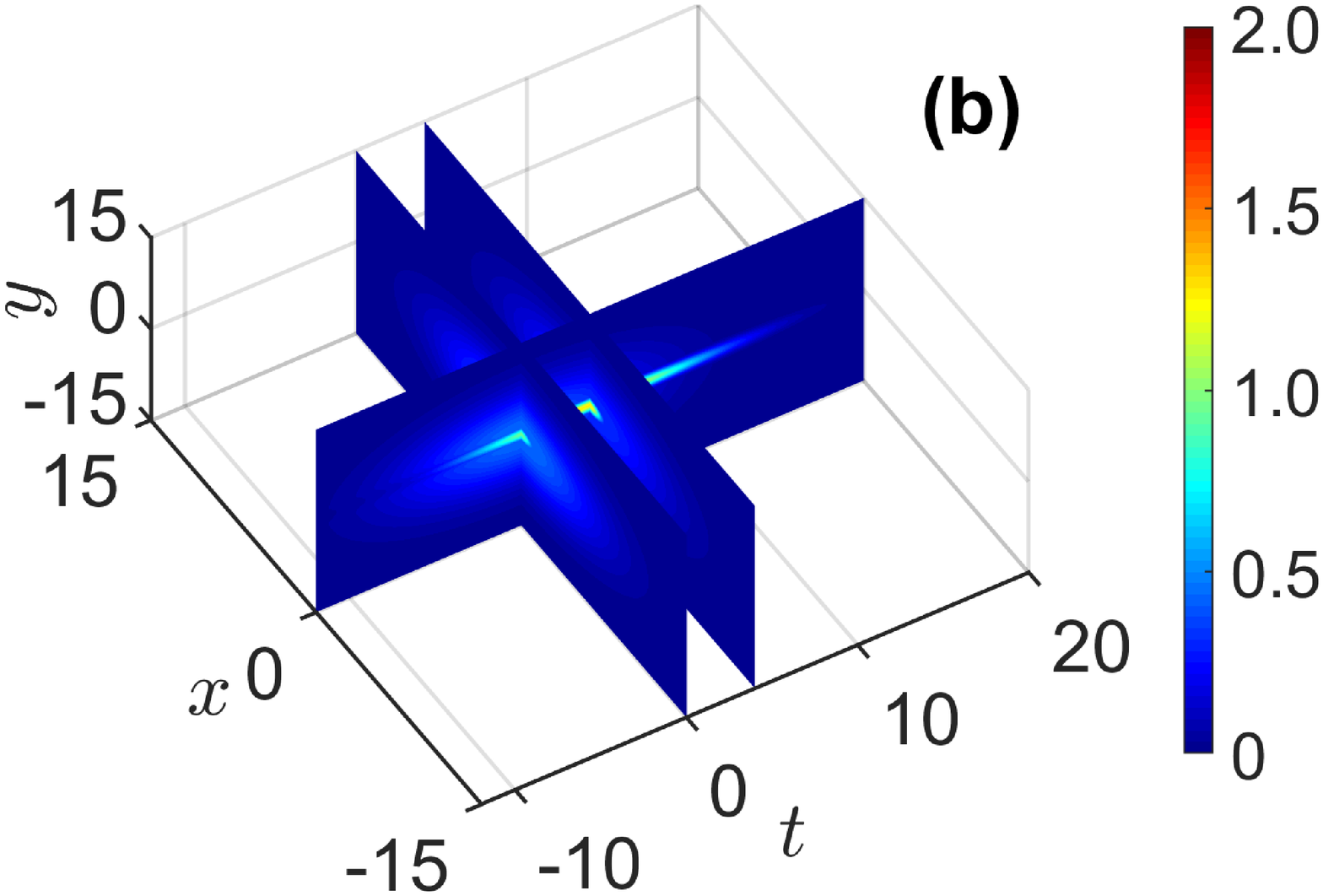}\\
\epsfxsize=10cm  \epsffile{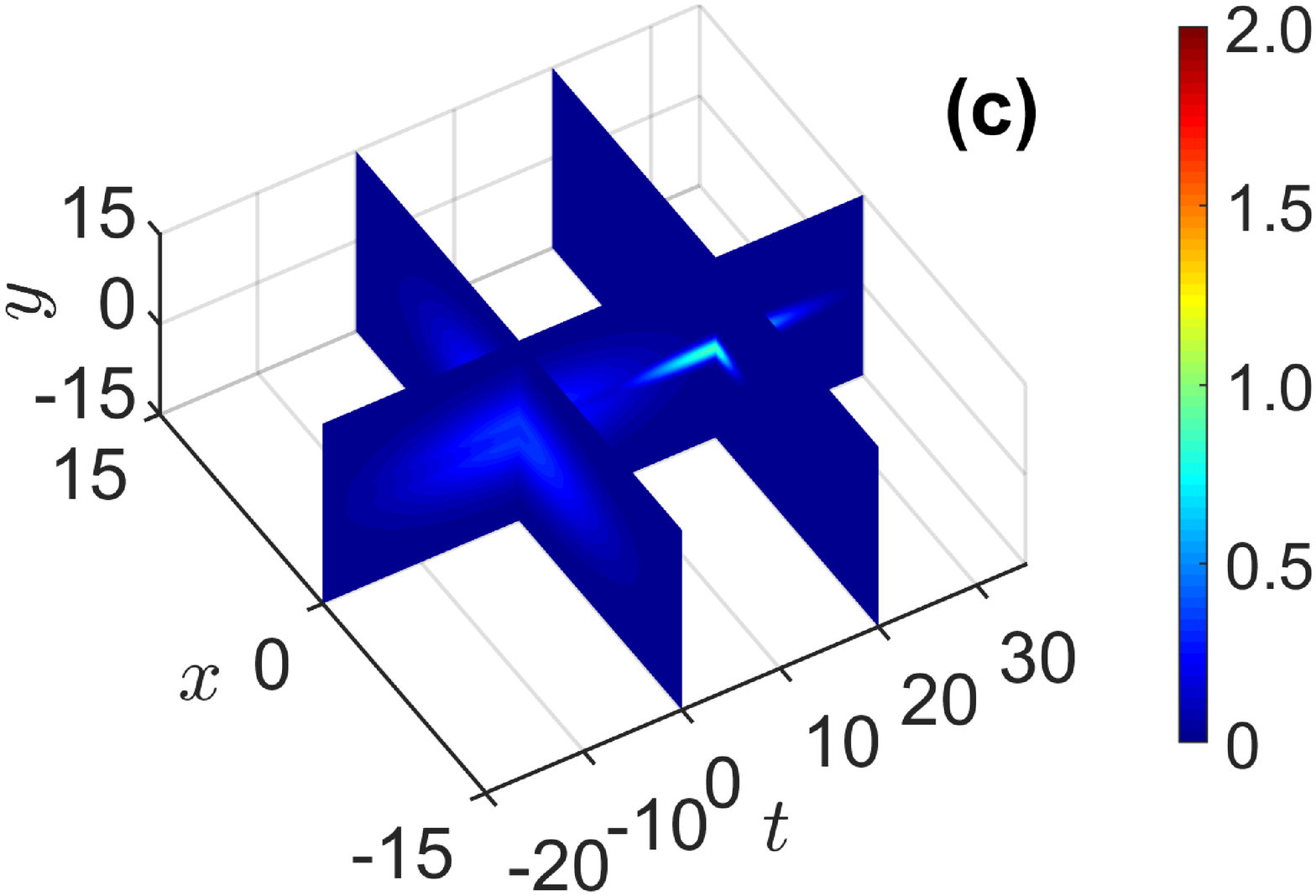}\\
\end{center}
%\caption{(Color online) The pulsed-beams 1 and 2 at $z=0$ (a), at $z=z_{i}=2.4$ (b), 
%and at $z=z_{f}=4$ (c) for $\Delta\beta_{1}=10$ and $\epsilon_3=0.12$.}
\caption{(Color online) 
Contour plots of the pulsed-beam shapes $|\psi_{j}(t,x,y,z)|$ 
on three planes in the $txy$ space at $z=0$ (a), $z=z_{i}=2.4$ (b), 
and $z=z_{f}=4.0$ (c) in a fast collision with strong spatially localized 
effects (collision setup 1). The cubic loss coefficient is $\epsilon_3=0.12$ 
and the first-order dispersion coefficient is $\Delta\beta_{1}=10$.   
The plots represent the pulsed-beam shapes obtained by numerical solution 
of Eq. (\ref{pb5}) with the initial condition (\ref{pb45}).
The three planes are $x=0$, $t=0$, and $t=t_{2}(z)$ with 
$z=0$, $z=z_{i}$, and $z=z_{f}$ in (a), (b), and (c), respectively.}
\label{fig1}
\end{figure}

\begin{figure}[ptb]
\begin{center}
\epsfxsize=16.5cm  \epsffile{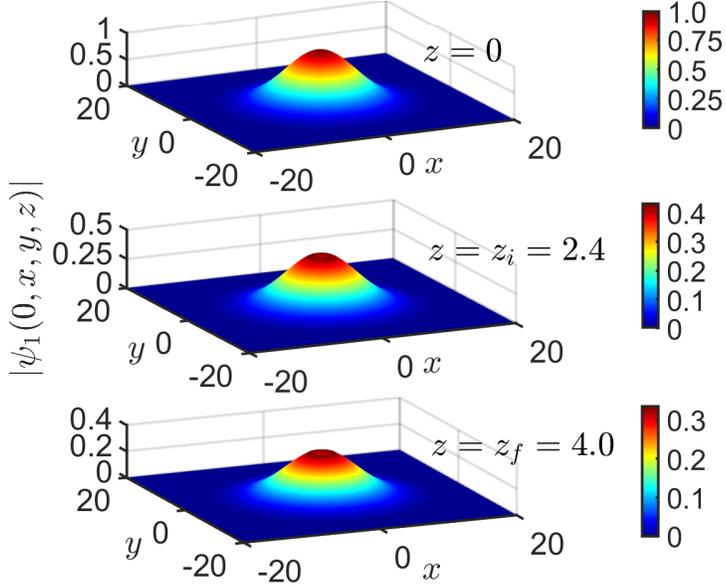}
\end{center}
\caption{(Color online) 
The shape of pulsed-beam 1 at $t=0$ $|\psi_{j}(0,x,y,z)|$ 
vs $x$ and $y$ at $z=0$ (top), $z=z_{i}=2.4$ (middle), and $z=z_{f}=4.0$ (bottom) 
in a fast two-beam collision with strong spatially localized effects. 
The parameter values are the same as in Fig. \ref{fig1}.   
The plots represent the shape of pulsed-beam 1 obtained by numerical 
solution of Eq. (\ref{pb5}) with the initial condition (\ref{pb45}).}
\label{fig2}
\end{figure}

We now turn to describe the results of the numerical simulation. 
Figure \ref{fig1} shows the pulsed-beam shapes $|\psi_{j}(t,x,y,z)|$ 
obtained in the simulation at three specific planes (cross-sections) 
at distance $z=0$, the intermediate distance $z=z_{i}=2.4$, and the 
final distance $z=z_{f}=4.0$ \cite{cross_sections,zi_values}. We observe 
that the pulsed-beams experience broadening due to the effects of 
second-order dispersion and diffraction. Furthermore, the values 
of $|\psi_{j}(t,x,y,z)|$ in the main bodies of the pulsed-beams 
decrease significantly. We further characterize the intensity decrease 
for pulsed-beam 1 by presenting the graphs of $|\psi_{1}(0,x,y,z)|$ 
vs $x$ and $y$ at $z=0$, $z=z_{i}$, and $z=z_{f}$ in Fig. \ref{fig2}. 
It is seen that the intensity decrease is a result of both 
dispersion-induced (and diffraction-induced) broadening and the 
relatively strong effect of cubic loss on the collision. 
Moreover, the intensity reduction is spatially localized 
near the $z$ axis, and it leads to the generation of a local 
minimum in the graph of $|\psi_{1}(0,x,y,z)|$ 
vs $x$ and $y$ on the $z$ axis.

% fig 2 - the shape of pulsed-beam 1 - setup 1   

\begin{figure}[ptb]
\begin{center}
\epsfxsize=15cm \epsffile{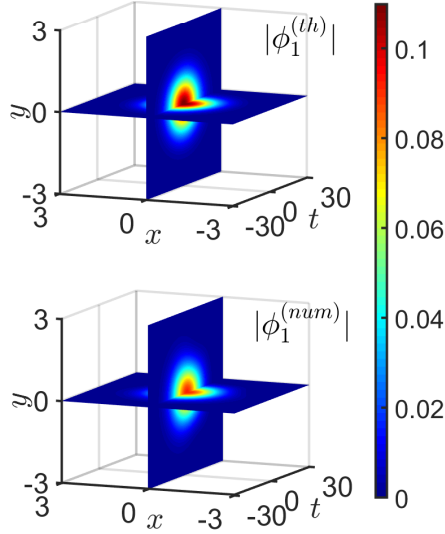}
\end{center}
\caption{(Color online) 
The collision-induced change in the shape of pulsed-beam 1 $|\phi_{1}(t,x,y,z_{f})|$ 
at $z_{f}=4.0$ in a fast two-beam collision with parameter values  
$\epsilon_{3}=0.12$ and $\Delta\beta_{1}=10$ (collision setup 1).   
Top: the perturbation theory prediction of Eq. (\ref{pb70}). 
Bottom: the result obtained by numerical solution of Eq. (\ref{pb5}).} 
\label{fig3}
\end{figure}

The validity of the perturbation theory prediction for the collision-induced 
change in the shape of pulsed-beam 1 $\phi_{j}(t,x,y,z)$ is checked 
in Fig. \ref{fig3}. More specifically, this figure shows a comparison 
between the theoretical prediction of Eq. (\ref{pb70}) $|\phi_{1}^{(th)}(t,x,y,z)|$ 
and the numerical simulation's result $|\phi_{1}^{(num)}(t,x,y,z)|$ at $z=z_{f}$. 
We observe that the magnitude of the collision-induced change in the pulsed-beam's 
shape is much larger than the one observed in the collision setups considered          
in Ref. \cite{PHN2022} for time-independent optical beams, and in Ref. \cite{PC2020} 
for conventional optical solitons in dimension 1. (Compare Fig. \ref{fig3} with Fig. 7 in Ref. 
\cite{PHN2022} and with Fig. 1 in Ref. \cite{PC2020}). Furthermore, despite the 
relatively strong collision-induced changes seen in Fig. \ref{fig3}, 
the agreement between the theoretical prediction and the simulation's 
result is good. We quantify the deviation of $|\phi_{1}^{(th)}(t,x,y,z)|$ 
from $|\phi_{1}^{(num)}(t,x,y,z)|$ by defining the relative error 
(in percentage) $E_{r}^{(|\phi_{1}|)}(z)$ as  
\begin{eqnarray}&&
E_{r}^{(|\phi_{1}|)}(z)=
100 \times
\left [\int dt \int dx \int dy \, |\phi_{1}^{(th)}(t,x,y,z)|^{2} \right ]^{-1/2}
\nonumber \\&&
\times
\left\{\int dt \int dx \int dy 
\left[\;\left|\phi_{1}^{(th)}(t,x,y,z) \right| - 
\left|\phi_{1}^{(num)}(t,x,y,z) \right| \; 
\right]^2 \right\}^{1/2},   
\label{pb75}
\end{eqnarray}               
where the integration is performed over the entire simulation domain. 
The calculated value of $E_{r}^{(|\phi_{1}|)}(z_{f})$ for the collision 
setup considered here is $17.5\%$, in accordance with the good agreement 
between theory and simulation  observed in Fig. \ref{fig3}.

% fig 3 - collision-induced change in the shape of pulsed-beam 1 
% setup 1  

\begin{figure}[ptb]
\begin{center}
\epsfxsize=12cm  \epsffile{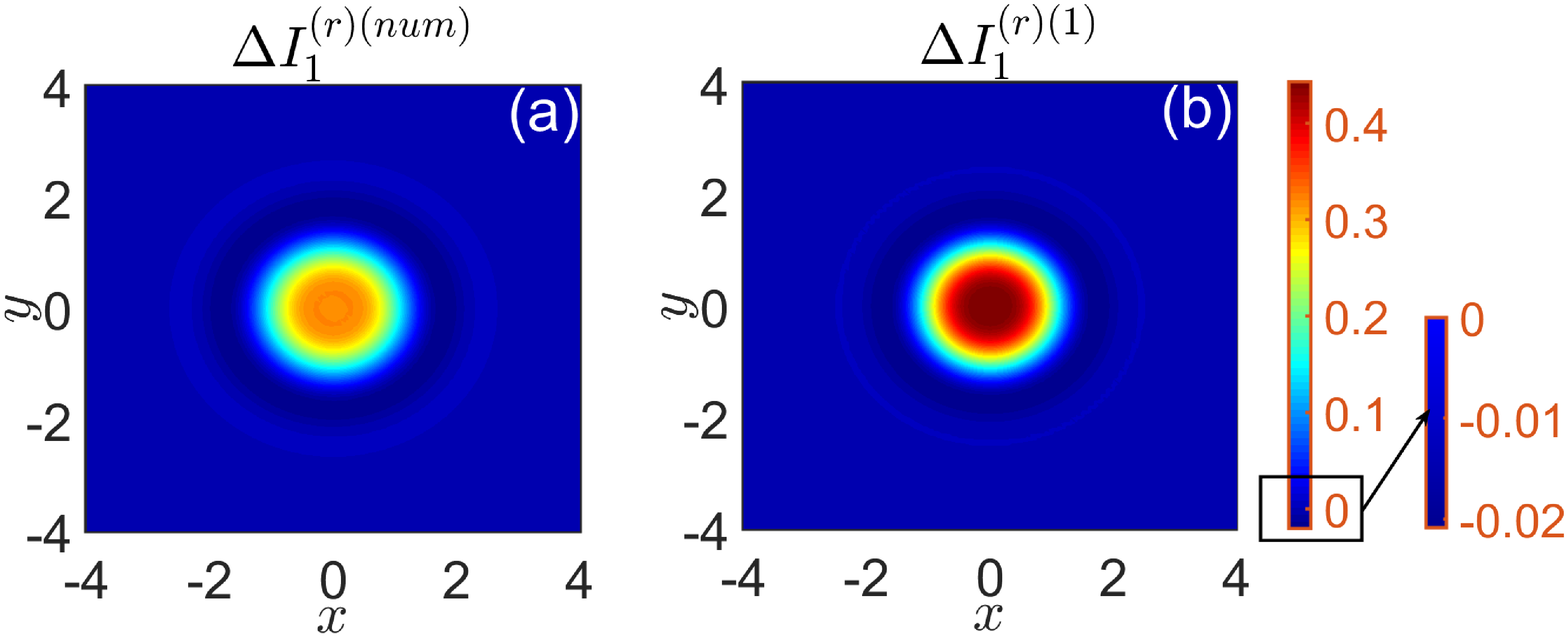}\\
\epsfxsize=12cm  \epsffile{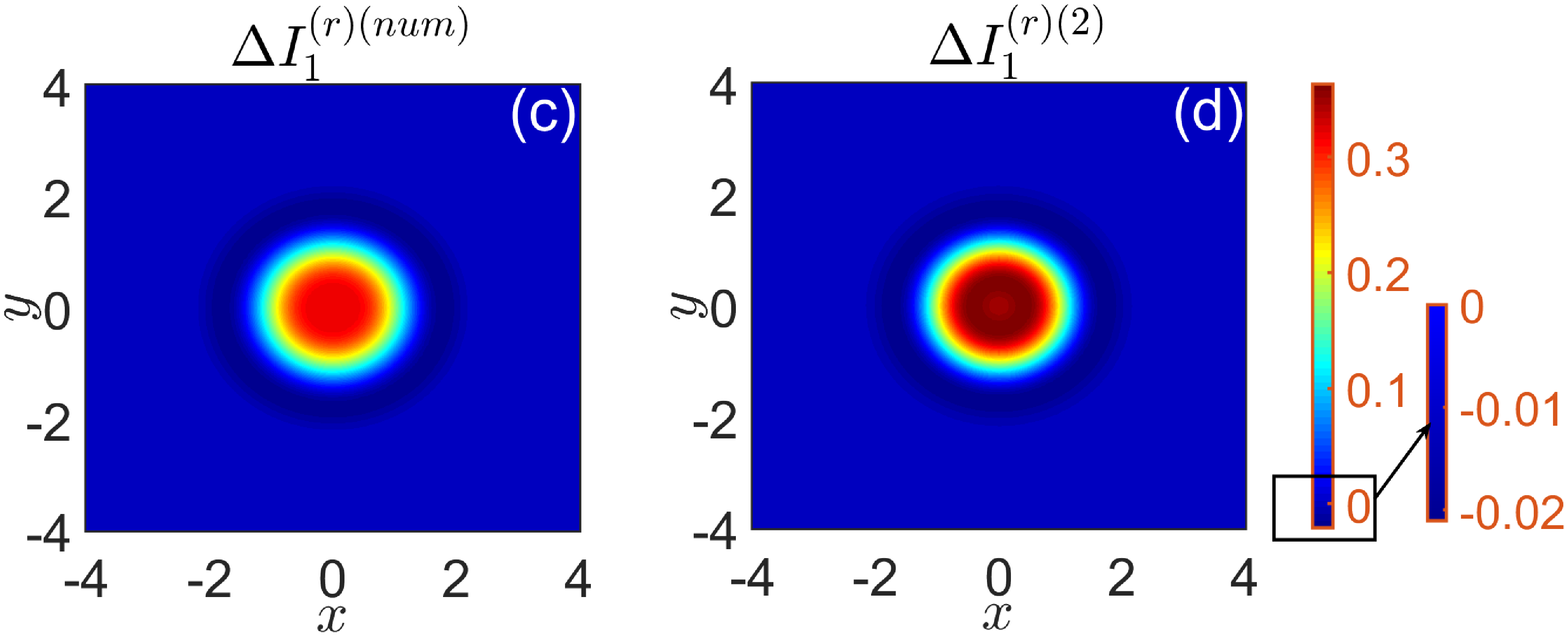}\\
\end{center}
\caption{(Color online) 
The fractional intensity reduction factor for pulsed-beam 1 $\Delta I_{1}^{(r)}(x,y,z)$ 
vs $x$ and $y$ at $z=z_{f}$ in a fast two-beam collision with parameter values  
$\epsilon_{3}=0.12$ and $\Delta\beta_{1}=10$ (collision setup 1). 
The result obtained in the simulation with Eq. (\ref{pb5}) is shown in (a) and (c). 
The perturbation theory predictions of Eqs. (\ref{pb49}) and (\ref{pb77}) are shown 
in (b) and (d), respectively.}           
\label{fig4}
\end{figure}

Important insight into the collision-induced effects is gained by 
analyzing the behavior of the fractional intensity reduction factor 
$\Delta I_{1}^{(r)}(x,y,z)$. The dependence of the intensity 
reduction factor on $x$ and $y$ at $z=z_{f}$ that was obtained in the simulation  
with Eq. (\ref{pb5}) $\Delta I_{1}^{(r)(num)}$ is shown in Fig. \ref{fig4}(a) \cite{Delta_I_1}. 
The perturbation theory prediction of Eq. (\ref{pb49}) $\Delta I_{1}^{(r)(1)}$ 
is shown in Fig. \ref{fig4}(b). We observe good agreement between the 
two results. In particular, the values of $\Delta I_{1}^{(r)(num)}$ 
are larger than 0.25 within a disk of radius $R \simeq 0.8$, which is much 
smaller than the initial pulsed-beam width $W_{10}^{(x)}=W_{10}^{(y)}=5$, 
in accordance with the perturbation theory prediction for strong localized 
intensity reduction effects. We also note that the perturbation theory prediction 
overestimates the values of the intensity reduction factor in the main body of the 
pulsed-beam. For example, the maximal values of $\Delta I_{1}^{(r)(1)}$ 
and $\Delta I_{1}^{(r)(num)}$, which are attained at $R \approx 0.1$ and 
$R \approx 0.3$, are 0.443 and 0.319, respectively. We further quantify the 
deviation of $\Delta I_{1}^{(r)(1)}$ from $\Delta I_{1}^{(r)(num)}$ 
by defining the relative error $E_{r}^{(\Delta I_{1}^{(r)(1)})}(z)$ (in percentage): 
\begin{eqnarray}&&
E_{r}^{(\Delta I_{1}^{(r)(1)})}(z)=
100 \times
\left [\int dx \int dy \, |\Delta I_{1}^{(r)(1)}(x,y,z)|^{2} \right ]^{-1/2}
\nonumber \\&&
\times
\left\{\int dx \int dy 
\left[\;\left| \Delta I_{1}^{(r)(1)}(x,y,z) \right| - 
\left|\Delta I_{1}^{(r)(num)}(x,y,z) \right| \; 
\right]^2 \right\}^{1/2}.   
\label{pb76}
\end{eqnarray}            
The value of $E_{r}^{(\Delta I_{1}^{(r)(1)})}(z_{f})$ for the current collision 
setup is $23.6\%$, which is consistent with the comparison 
shown in Fig. \ref{fig4}.

% fig 4 the fractional intensity reduction factor vs x and y - setup 1 

The deviation of $\Delta I_{1}^{(r)(1)}$ from $\Delta I_{1}^{(r)(num)}$ 
is clearly associated with the strong effects of the collision. 
Therefore, one can expect that better agreement between theory and 
simulation would be obtained by including effects of order 
higher than $\epsilon_{3}/|\Delta\beta_{1}|$ 
in the perturbative calculation of $\Delta I_{1}^{(r)}$. 
In the current paper we do not carry out the full calculation of 
high-order contributions to $\Delta I_{1}^{(r)}$, since as we 
will see below, this lengthy calculation is not essential for 
obtaining significant improvement in the agreement between theory and simulation. 
Instead, we add only the correction to $\Delta I_{1}^{(r)}$ due to the 
$O(\epsilon_{3}^{2}/|\Delta\beta_{1}|^{2})$ contribution from the term 
$-|\phi_{1}|^{2}/(A_{1}^{2}(0) \Psi_{10}^{2})$. Consequently, the improved 
prediction for the intensity reduction factor $\Delta I_{1}^{(r)(2)}$ 
is given by:             
\begin{eqnarray}&&
\!\!\!\!\!\!\!\!\!\!\!\!\!\!
\Delta I_{1}^{(r)(2)}(x,y,z)= 
-\frac{2|\phi_{1}(t,x,y,z)|\cos\left[\Delta \chi_{1}^{(tot)}(t,x,y,z)\right]}
{A_{1}(0) \Psi_{10}(t,x,y,z)}
%\nonumber \\&&
-\frac{|\phi_{1}(t,x,y,z)|^{2}}{A_{1}^{2}(0)
\Psi_{10}^{2}(t,x,y,z)}.
\label{pb77}
\end{eqnarray}
The comparison between $\Delta I_{1}^{(r)(2)}$ and $\Delta I_{1}^{(r)(num)}$ 
is shown in the bottom row of Fig. \ref{fig4}. It is seen that the agreement 
between the two results is significantly better than the agreement between 
$\Delta I_{1}^{(r)(1)}$ and $\Delta I_{1}^{(r)(num)}$. Accordingly, the relative 
error (in percentage) in the approximation of the intensity reduction factor 
by $\Delta I_{1}^{(r)(2)}$, $E_{r}^{(\Delta I_{1}^{(r)(2)})}(z)$, 
which is defined by an equation similar to Eq. (\ref{pb76}), 
is only $13.4\%$ at $z=z_{f}$. Thus, the introduction of the 
term $-|\phi_{1}|^{2}/(A_{1}^{2}(0) \Psi_{10}^{2})$ does lead 
to significant improvement in the accuracy of the perturbation theory 
prediction. It follows that the relatively strong intensity reduction 
effects observed in the current collision setup can be correctly 
explained by using only the leading-order term in the expansion of 
$\phi_{1}$ (i.e., the term of order $\epsilon_{3}/|\Delta\beta_{1}|$). 
This finding is important since it means that the design of the 
collision setups that are considered in the current paper, 
which lead to significant intensity reduction effects,   
can indeed be based on the relatively simple calculation of $\phi_{1}$ 
in the leading order of the perturbative calculation.

\begin{figure}[ptb]
\begin{center}
\epsfxsize=12cm \epsffile{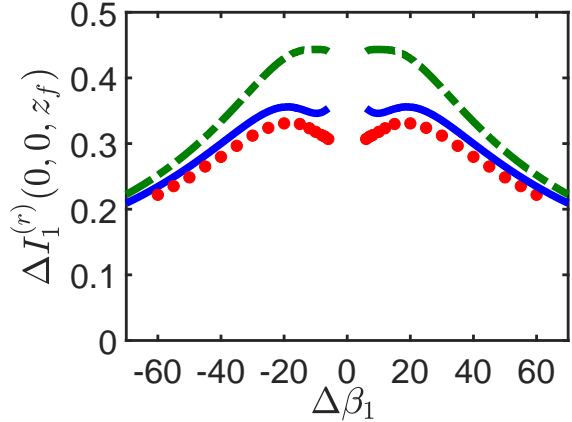}
\end{center}
\caption{(Color online) 
The fractional intensity reduction factor for pulsed-beam 1 at $(x,y)=(0,0)$ 
and $z=z_{f}$, $\Delta I_{1}^{(r)}(0,0,z_{f})$, vs the first-order dispersion 
coefficient $\Delta\beta_{1}$ in fast two-beam collisions with $\epsilon_{3}=0.12$ 
(collision setup 1). The red circles represent the result obtained by 
numerical simulations with Eq. (\ref{pb5}). The dashed-dotted green and 
solid blue curves correspond to the perturbation theory predictions 
of Eqs. (\ref{pb49}) and (\ref{pb77}), respectively.}  
\label{fig5}
\end{figure}

To gain further understanding of the effects of cubic loss on fast 
collisions we investigate the dependences of 
$\Delta I_{1}^{(r)}(0,0,z_{f})$ (the value of the intensity reduction 
factor on the $z$ axis) and $\Delta A_{1}^{(c)}$ (the collision-induced 
amplitude shift) on the first-order dispersion coefficient $\Delta\beta_{1}$. 
For this purpose, we carry out extensive numerical simulations with 
Eq. (\ref{pb5}) with the parameter values specified in the beginning of 
the current subsection and with $\Delta\beta_{1}$ values in the intervals 
$6 \le |\Delta\beta_{1}| \le 60$. We denote the values of 
$\Delta I_{1}^{(r)}(0,0,z_{f})$ and $\Delta A_{1}^{(c)}$ obtained 
in the simulations by $\Delta I_{1}^{(r)(num)}(0,0,z_{f})$ and 
$\Delta A_{1}^{(c)(num)}$, respectively. Figure \ref{fig5} shows 
the dependence of $\Delta I_{1}^{(r)(num)}(0,0,z_{f})$ on $\Delta\beta_{1}$ 
along with the perturbation theory predictions of Eqs. (\ref{pb49}) 
and (\ref{pb77}), $\Delta I_{1}^{(r)(1)}(0,0,z_{f})$ and 
$\Delta I_{1}^{(r)(2)}(0,0,z_{f})$. We first note that the values of     
$\Delta I_{1}^{(r)(num)}(0,0,z_{f})$ are larger than 0.2 over the 
entire $\Delta\beta_{1}$ intervals, in accordance with the perturbation 
theory predictions for strong intensity reduction. We also note that 
these values are larger by two orders of magnitude compared with the 
values obtained in Ref. \cite{PHN2022} for fast collisions between 
time-independent beams. Another interesting feature seen in Fig. \ref{fig5} 
is that $\Delta I_{1}^{(r)(num)}(0,0,z_{f})$ has two local maxima at 
$\Delta\beta_{1} \approx \pm 20.0$. These local maxima are correctly captured 
by the perturbation theory prediction $\Delta I_{1}^{(r)(2)}$ but not by 
$\Delta I_{1}^{(r)(1)}$. More generally, we observe good agreement between 
$\Delta I_{1}^{(r)(1)}(0,0,z_{f})$ and $\Delta I_{1}^{(r)(num)}(0,0,z_{f})$ 
and very good agreement between $\Delta I_{1}^{(r)(2)}(0,0,z_{f})$ 
and $\Delta I_{1}^{(r)(num)}(0,0,z_{f})$ over the entire $\Delta\beta_{1}$ 
intervals. Importantly, the agreement between the two theoretical 
results and the simulations result improves with increasing 
value of $|\Delta\beta_{1}|$. Therefore, the collision setup considered 
in the current subsection leads to significant intensity reduction 
effects not only for intermediate values of $|\Delta\beta_{1}|$, 
but also for large $|\Delta\beta_{1}|$ values. 
Note that the collision distance $z_{c}$ is inversely proportional 
to  $|\Delta\beta_{1}|$. Therefore, the latter finding means that 
significant intensity reduction effects can also be observed at 
much shorter distances than the dimensional collision distance $z''_{c}=216.0$ m 
of the collision setup with $\Delta\beta_{1}=10$. For example, 
for $\Delta\beta_{1}=50$, the dimensional collision distance is 
only $z''_{c}=43.2$ m, and the value of 
$\Delta I_{1}^{(r)(num)}(0,0,z_{f})$ is 0.248.

% fig 5 - the \Delta\beta_{1} dependence of the fractional intensity 
% reduction factor for pulsed-beam 1 at (x,y)=(0,0) and z=z_{f}. 
% collision setup 1  

\begin{figure}[ptb]
\begin{center}
\epsfxsize=12cm \epsffile{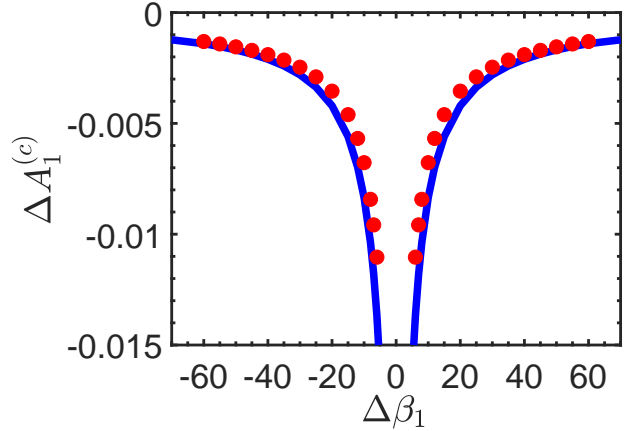}
\end{center}
\caption{(Color online) 
The collision-induced amplitude shift of pulsed-beam 1 $\Delta A_{1}^{(c)}$ 
vs the first-order dispersion coefficient $\Delta\beta_1$ in fast two-beam 
collisions with $\epsilon_3=0.12$ (collision setup 1). The red circles represent   
the result obtained by numerical simulations with Eq. (\ref{pb5}).  
The solid blue curve represents the perturbation theory prediction  
of Eq. (\ref{pb71}).}  
\label{fig6}
\end{figure}

The dependence of the collision-induced amplitude shift 
$\Delta A_{1}^{(c)}$ on $\Delta\beta_1$ that is obtained 
in the simulations is shown in Fig. \ref{fig6} 
together with the perturbation theory prediction of Eq. (\ref{pb71}). 
We observe good agreement between the perturbation theory and 
the simulations over the entire $\Delta\beta_1$ intervals 
despite the strong intensity reduction effects. More specifically, 
the relative error in the approximation of $\Delta A_{1}^{(c)}$ 
(in percentage), which is defined by 
$E_{r}=100|\Delta A_{1}^{(c)(num)}-\Delta A_{1}^{(c)(th)}|
/|\Delta A_{1}^{(c)(th)}|$, is smaller than 
$19.4\%$ for $6\le |\Delta\beta_1| <30$ and smaller than 
$12.2\%$ for $30\le |\Delta\beta_1| \le 60$. 
Therefore, the perturbative calculation that is used to 
design the collision setups that lead to strong intensity 
reduction effects also correctly captures the behavior of the 
collision-induced amplitude shift.  
We also observe that the values of $\Delta A_{1}^{(c)(num)}$ 
in Fig. \ref{fig6} are smaller by one to two 
orders of magnitude compared with the values of 
$\Delta I_{1}^{(r)(num)}(0,0,z_{f})$ in Fig. \ref{fig5}. 
This can be explained by noting that $\Delta I_{1}^{(r)}$ is 
a measure for localized intensity changes. In contrast, 
$\Delta A_{1}^{(c)}$ is a measure for global (total) intensity 
changes, since its calculation involves integration over the 
spatial coordinates [see Eqs. (\ref{pb23}) and (\ref{pb34})].        
In the current setup, the strong intensity reduction effects 
are localized in a small region near the $z$ axis (see the discussion 
of Fig. \ref{fig4}). As a result, this strong but spatially 
localized intensity reduction leads only to relatively small 
amplitude shifts. In the next subsection, we consider a collision 
setup with nonlocalized intensity reduction effects, for which 
the values of both $\Delta A_{1}^{(c)(num)}$ and 
$\Delta I_{1}^{(r)(num)}(0,0,z_{f})$ are relatively large.

% fig 6 - \Delta A_{1}^{(c)} vs \Delta\beta_1 - setup 1 

\subsection{A collision setup with nonlocalized effects}
\label{nonlocalized}

We now turn to discuss collision setup 2, which leads to strong but nonlocalized 
intensity reduction effects. It is interesting to consider this second setup,    
since some important features of the collision-induced effects in this case are 
very different from the ones observed for collision setup 1. 
The values of the physical parameters for setup 2 are chosen by 
using the conditions (\ref{pb54}), (\ref{pb55}), 
and (\ref{pb60}) in the following manner.

\begin{enumerate}

\item Nonlocalized intensity reduction for pulsed-beam 1 
is realized by choosing $W_{10}^{(x)}=W_{10}^{(y)}=3$ and  
$W_{20}^{(x)}=W_{20}^{(y)}=2$, in accordance with condition (\ref{pb60}). 

\item We choose $t_{10}=0$ and $t_{20}=-20$, such that the pulsed-beams 
are well-separated at $z=0$. We also choose $\Delta\beta_{1}=20$, 
and therefore, $z_{c}=1$. The value of $d_{21}$ is chosen as 1.0, 
and therefore, the diffraction-induced broadening of pulsed-beam 1 
is nonnegligible at $z=z_{c}$. 

\item Since $\lambda_{1}$ is close to $\lambda_{2}$, it follows 
from the relation $d_{22}=\lambda_{2}d_{21}/\lambda_{1}$ that 
$d_{22} \approx d_{21}$. We therefore use $d_{22}=1.0$.
As a result, the diffraction-induced broadening of pulsed-beam 2 
at $z=z_{c}$ is also nonnegligible.  
   
\item Strong intensity reduction effects are realized by choosing 
$\epsilon_{3}=0.25$, $A_{2}(0)=1.3$, and $W_{20}^{(t)}=5$, 
in conformity with Eqs. (\ref{pb54}) and (\ref{pb55}).   
This choice yields $\Delta I_{1}^{(r)(1)}(0,0,z_{c}^{+}) \simeq 0.599$.    

\item The other physical parameters values are taken as 
$\beta_{22}=1$, $A_{1}(0)=1$, and $W_{10}^{(t)}=1$. In addition, 
$z_{f}=2z_{c}=2$, such that the pulsed-beams are also well-separated 
at $z=z_{f}$.  

\end{enumerate}

It is useful to provide an example for the values of the dimensional 
physical parameters that correspond to the values of the 
dimensionless parameters in items 1-5. Considering bulk fused silica 
as an example for the optical medium and using the information 
in Refs. \cite{Malitson65,Tan98}, we find the following values. 
$x^{\prime}_{0}=2$ cm, $\lambda_{1}=2.791$ $\mu$m, 
$\lambda_{2}=2.795$ $\mu$m, $\tilde \beta_{21}=-394.23$ 
$\mbox{ps}^{2}\mbox{km}^{-1}$, $\tilde \beta_{22}=-396.59$ 
$\mbox{ps}^{2}\mbox{km}^{-1}$, $\tilde d_{21}=2.221\times 10^{-7}$m, 
$\tilde d_{22}=2.224\times 10^{-7}$m, $\tau_{0}=9.421$ ps, 
$P_{0}=0.05$ W, $\rho_{3}=11.104$ $\mbox{W}^{-1}\mbox{km}^{-1}$,        
and $z''_{c}=450.2$ m \cite{parameters2}.

% fig 7 - pulsed-beams shapes - setup 2 
\begin{figure}[ptb]
\begin{center}
\epsfxsize=10cm  \epsffile{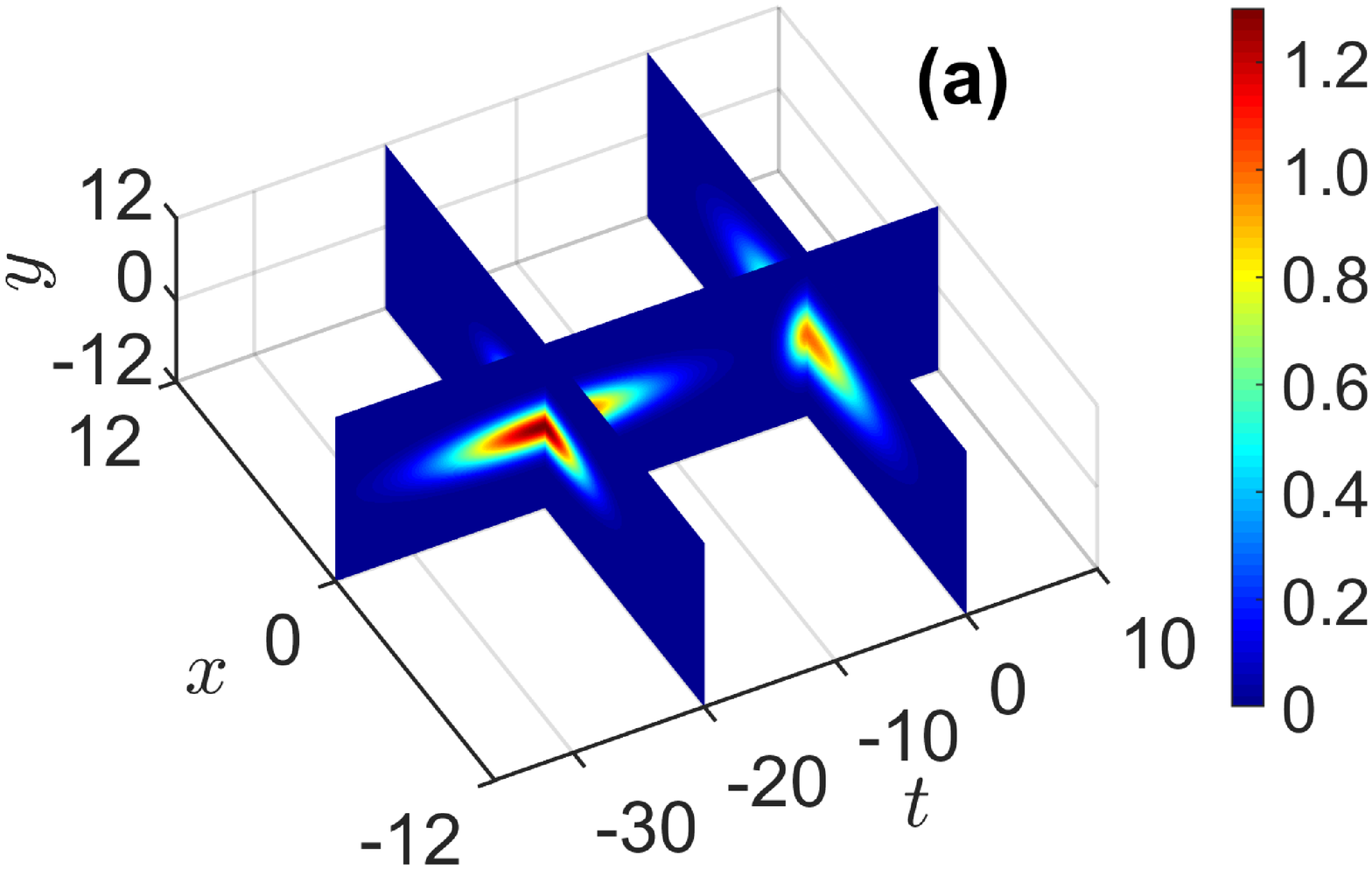}\\
\epsfxsize=10cm  \epsffile{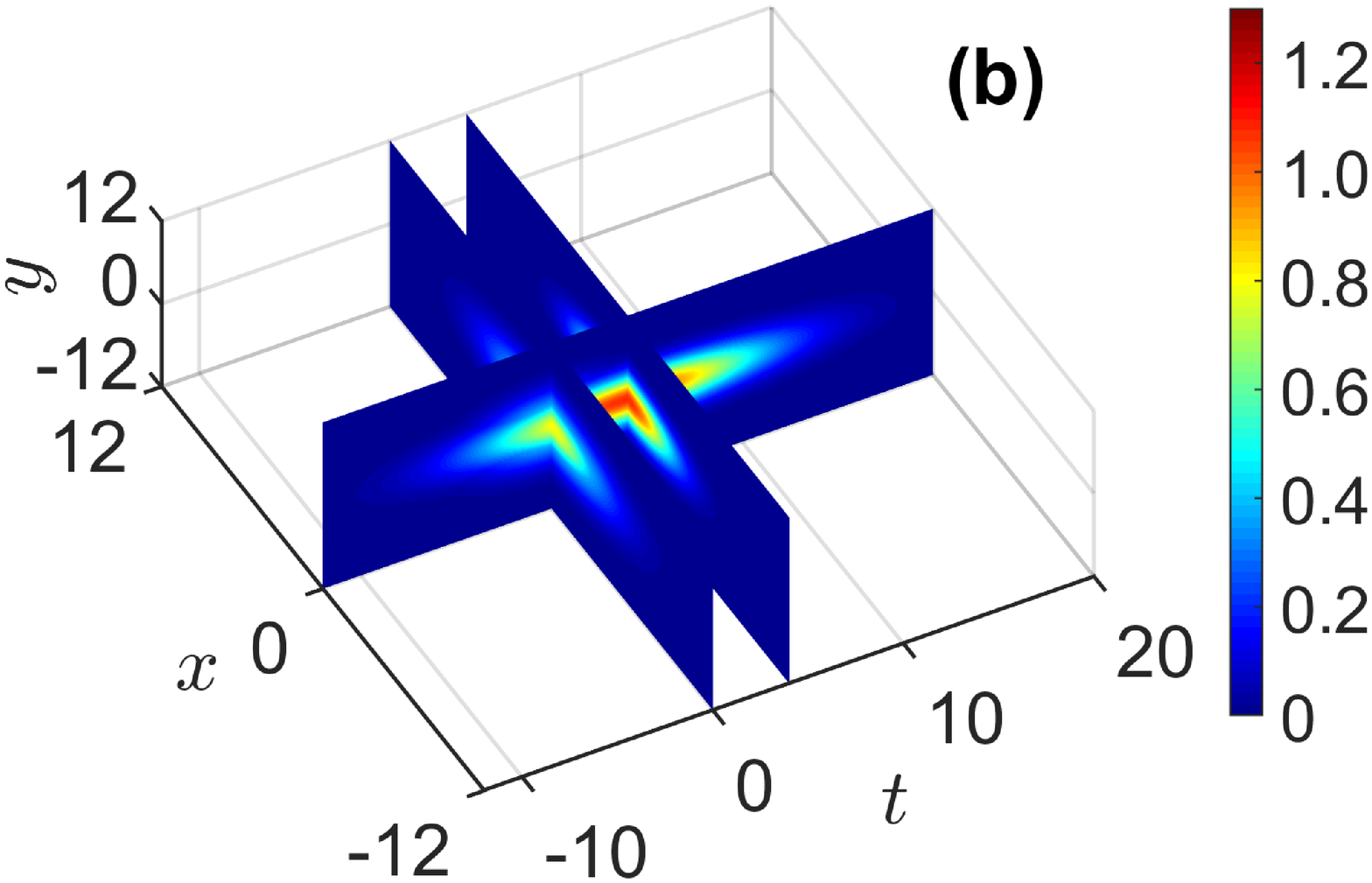}\\
\epsfxsize=10cm  \epsffile{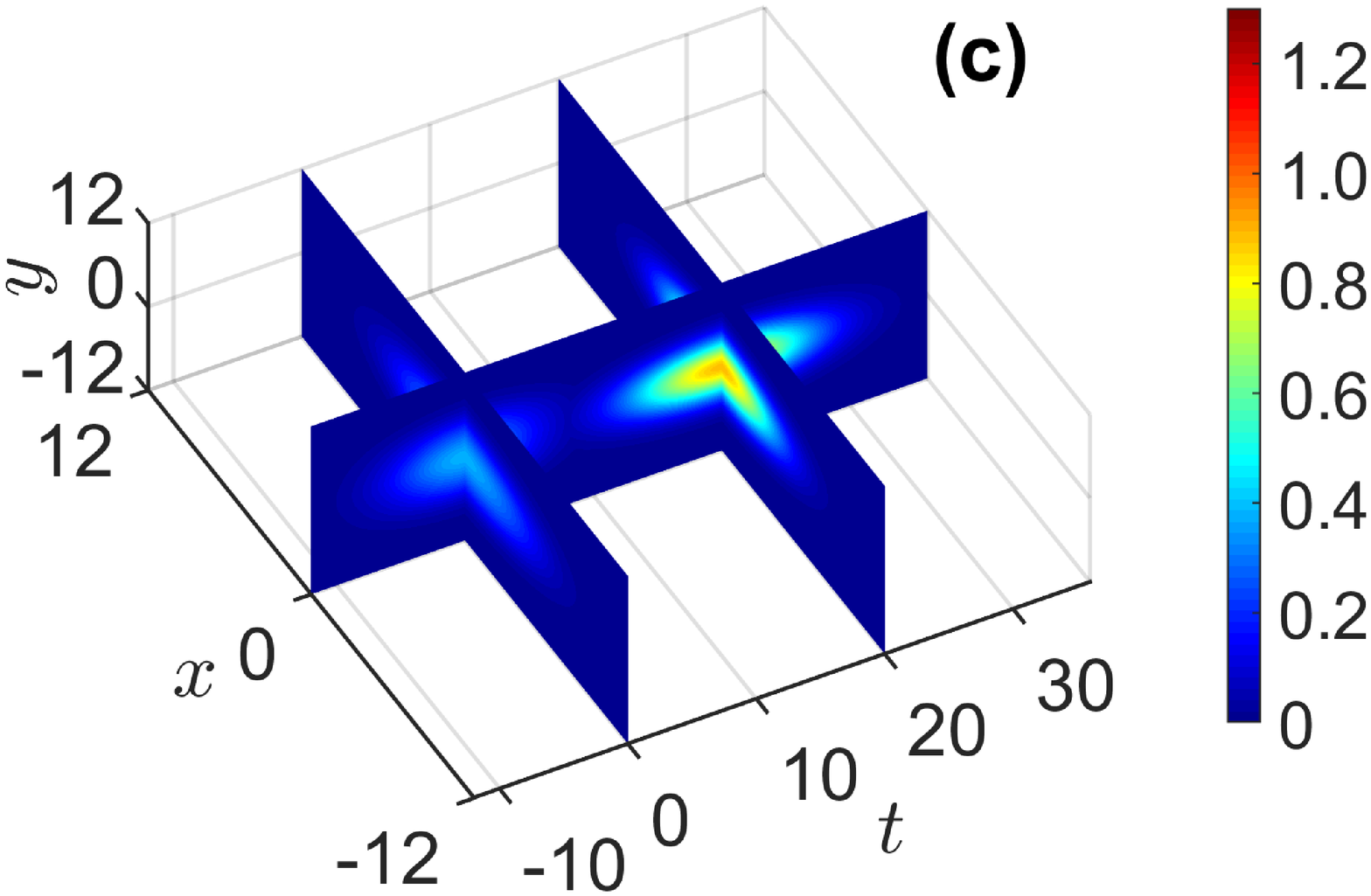}\\
\end{center}
\caption{(Color online) 
Contour plots of the pulsed-beam shapes $|\psi_{j}(t,x,y,z)|$ 
on three planes in the $txy$ space at $z=0$ (a), $z=z_{i}=1.2$ (b), 
and $z=z_{f}=2.0$ (c) in a fast collision with strong but nonlocalized 
effects (collision setup 2). The cubic loss coefficient is $\epsilon_3=0.25$ 
and the first-order dispersion coefficient is $\Delta\beta_{1}=20$.   
The plots represent the pulsed-beam shapes obtained by numerical solution 
of Eq. (\ref{pb5}) with the initial condition (\ref{pb45}).
The three planes are $x=0$, $t=0$, and $t=t_{2}(z)$ with 
$z=0$, $z=z_{i}$, and $z=z_{f}$ in (a), (b), and (c), respectively.}
\label{fig7}
\end{figure}

\begin{figure}[ptb]
\begin{center}
\epsfxsize=16.5cm  \epsffile{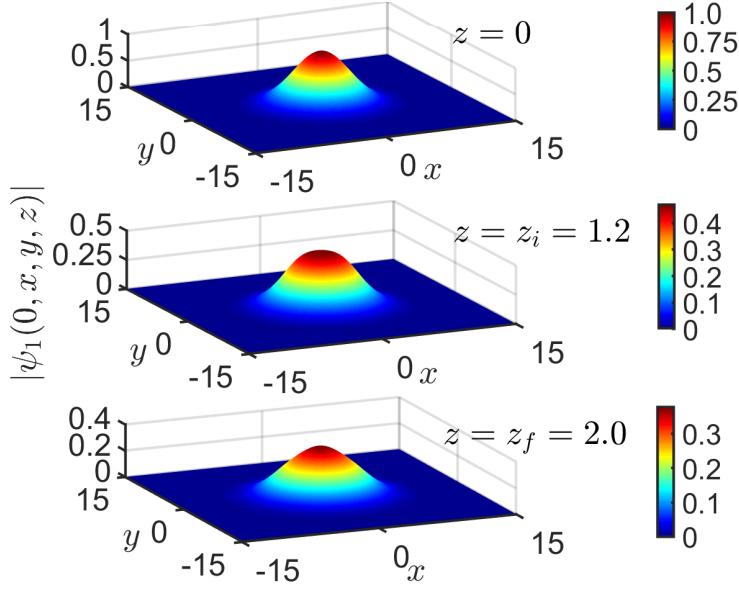}
\end{center}
\caption{(Color online) 
The shape of pulsed-beam 1 at $t=0$ $|\psi_{j}(0,x,y,z)|$ 
vs $x$ and $y$ at $z=0$ (top), $z=z_{i}=1.2$ (middle), and $z=z_{f}=2.0$ (bottom) 
in a fast two-beam collision with strong but nonlocalized effects. 
The parameter values are the same as in Fig. \ref{fig7}.   
The plots represent the shape of pulsed-beam 1 obtained by the numerical 
simulation with Eq. (\ref{pb5}) with the initial condition (\ref{pb45}).}
\label{fig8}
\end{figure}

Let us describe the results of the numerical simulation for collision setup 2. 
Figure \ref{fig7} shows the pulsed-beam shapes $|\psi_{j}(t,x,y,z)|$ 
obtained in the simulation at three specific planes at the distances 
$z=0$, $z=z_{i}=1.2$, and $z=z_{f}=2.0$. It is seen that the pulsed-beams 
broaden due to both second-order dispersion and diffraction. 
Additionally, the values of $|\psi_{j}(t,x,y,z)|$ in the main bodies 
of the pulsed-beams decrease significantly. The intensity decrease 
of pulsed-beam 1 is further characterized in Fig. \ref{fig8}, which 
shows the graphs of $|\psi_{1}(0,x,y,z)|$ vs $x$ and $y$ at $z=0$, 
$z=z_{i}$, and $z=z_{f}$. We observe that the intensity decrease of 
pulsed-beam 1 is a result of the interplay between 
dispersion-induced and diffraction-induced beam broadening 
and the strong effect of cubic loss on the collision. 
Furthermore, in contrast to the situation in setup 1, 
the intensity decrease is not localized in a small region near the 
$z$ axis, but affects the whole main body of the pulsed-beam.

% fig 8 - the shape of pulsed-beam 1 - setup 2   

\begin{figure}[ptb]
\begin{center}
\epsfxsize=15cm \epsffile{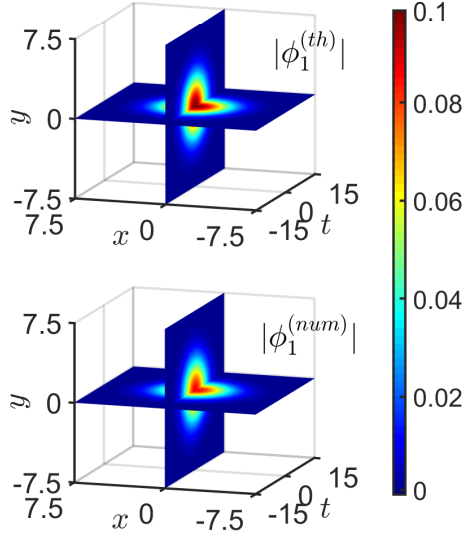}
\end{center}
\caption{(Color online) 
The collision-induced change in the shape of pulsed-beam 1 $|\phi_{1}(t,x,y,z_{f})|$ 
at $z_{f}=2.0$ in a fast two-beam collision with parameter values  
$\epsilon_{3}=0.25$ and $\Delta\beta_{1}=20$ (collision setup 2).   
Top: the perturbation theory prediction of Eq. (\ref{pb70}). 
Bottom: the result obtained by numerical solution of Eq. (\ref{pb5}).} 
\label{fig9}
\end{figure}

The nonlocalized character of the collisional effects is also 
seen in Fig. \ref{fig9}, which shows a comparison between the theoretical 
and numerical results for the collision-induced change in the shape 
of pulsed-beam 1 at $z=z_{f}$, $|\phi_{1}^{(th)}(t,x,y,z_{f})|$ and 
$|\phi_{1}^{(num)}(t,x,y,z_{f})|$. More specifically, the values of 
$|\phi_{1}^{(num)}(t,x,y,z_{f})|$ in Fig. \ref{fig9} are larger than 0.07 
in a ball of radius $R \approx 1.3$, whereas the $|\phi_{1}^{(num)}(t,x,y,z_{f})|$ 
values in Fig. \ref{fig3} are larger than 0.07 in a ball whose radius is only $R \approx 0.5$. 
Moreover, we find that the agreement between the theoretical prediction 
and the numerical simulation's result for $\phi_{1}$ in setup 2 is good 
despite the relatively strong effects of the collision. In particular, 
the relative error $E_{r}^{(|\phi_{1}|)}(z_{f})$, which is calculated 
with Eq. (\ref{pb75}), is only $13.2\%$.   
We also observe that the maximal value of $|\phi_{1}^{(num)}(t,x,y,z_{f})|$ 
in setup 2 is of the same order of magnitude as in setup 1.        
Thus, the magnitude of the collision-induced change in the pulsed-beam's 
shape in setup 2 is also much larger than the one observed in Refs. 
\cite{PHN2022} and \cite{PC2020} for fast collisions between 
time-independent optical beams and for fast collisions between 
optical solitons in dimension 1, respectively.

% fig 9 - collision-induced change in the shape of pulsed-beam 1 - setup 2  

\begin{figure}[ptb]
\begin{center}
\epsfxsize=12cm  \epsffile{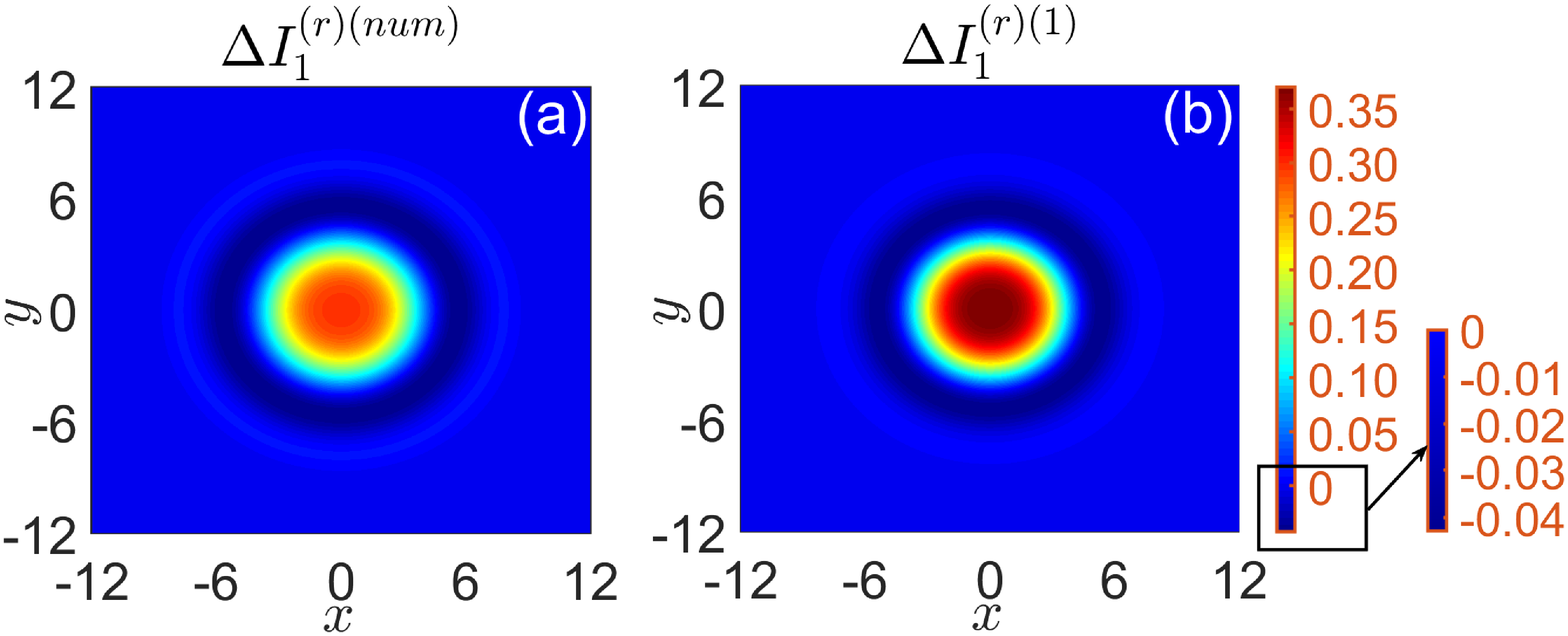}\\
\epsfxsize=12cm  \epsffile{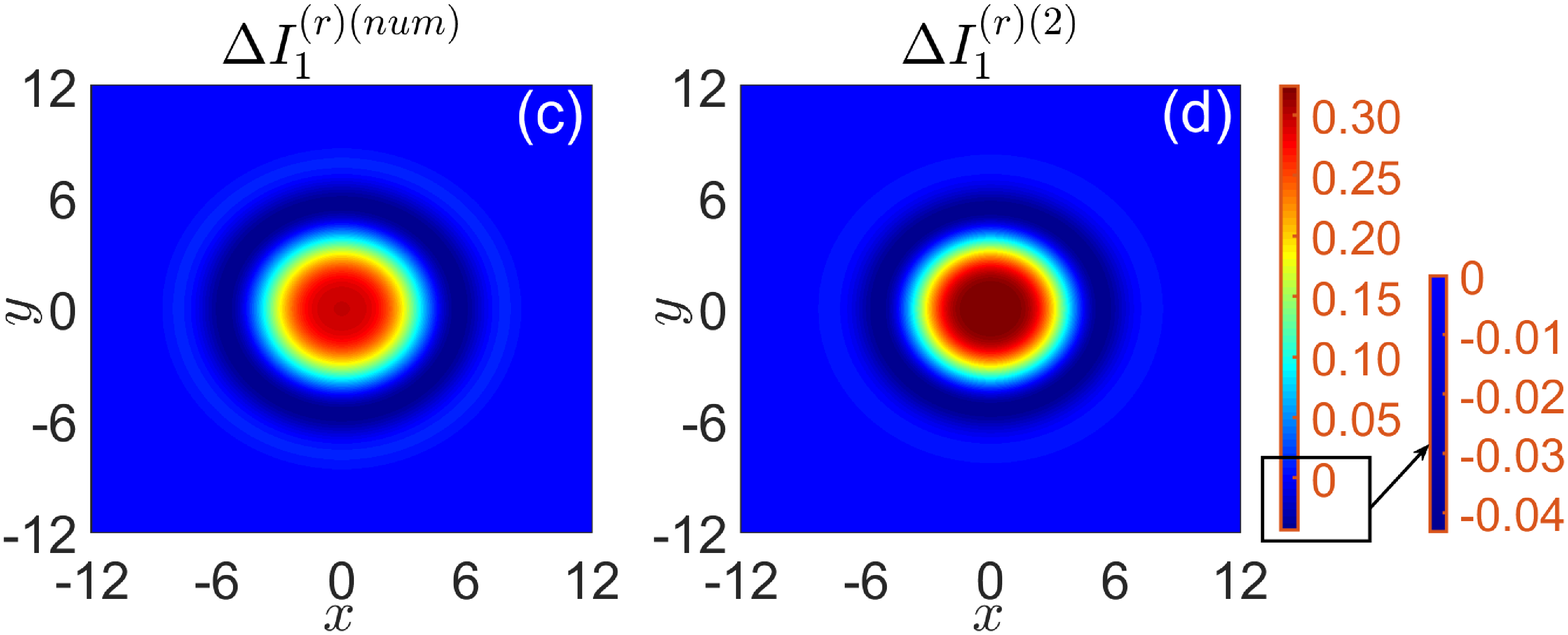}\\
\end{center}
\caption{(Color online) 
The fractional intensity reduction factor for pulsed-beam 1 $\Delta I_{1}^{(r)}(x,y,z)$ 
vs $x$ and $y$ at $z=z_{f}$ in a fast two-beam collision with parameter values  
$\epsilon_{3}=0.25$ and $\Delta\beta_{1}=20$ (collision setup 2). 
The result obtained by numerical solution of Eq. (\ref{pb5}) is shown in (a) and (c). 
The perturbation theory predictions of Eqs. (\ref{pb49}) and (\ref{pb77}) are shown 
in (b) and (d), respectively.}           
\label{fig10}
\end{figure}

To gain further insight into the collision dynamics we analyze the behavior 
of the fractional intensity reduction factor. Figure \ref{fig10}(a) shows 
the dependence of $\Delta I_{1}^{(r)(num)}$ on $x$ and $y$ at $z=z_{f}$ \cite{Delta_I_1}, 
and Fig. \ref{fig10}(b) shows the perturbation theory prediction of 
Eq. (\ref{pb49}) $\Delta I_{1}^{(r)(1)}$. We observe that unlike the 
situation in setup 1, the intensity reduction in setup 2 is nonlocalized 
and affects the entire main body of pulsed-beam 1. More specifically, 
the values of $\Delta I_{1}^{(r)(num)}$ are larger than 0.25 inside a 
disk of radius $R \approx 2.2$, which is close to the initial pulsed-beam 
width $W_{10}^{(x)}=W_{10}^{(y)}=3$, and larger than the value obtained 
in setup 1. We also observe good agreement between $\Delta I_{1}^{(r)(1)}$ 
and $\Delta I_{1}^{(r)(num)}$. Indeed, the value of the relative error 
$E_{r}^{(\Delta I_{1}^{(r)(1)})}(z_{f})$ that is calculated with Eq. (\ref{pb76}) 
is $18.6\%$, which is smaller than the value obtained in setup 1. 
Further improvement in the agreement between the theoretical and the numerical 
results is obtained by employing the perturbation theory prediction 
$\Delta I_{1}^{(r)(2)}$, which is defined by Eq. (\ref{pb77}). 
The comparison between $\Delta I_{1}^{(r)(2)}$ and $\Delta I_{1}^{(r)(num)}$ 
is shown in Figs. \ref{fig10}(c) and \ref{fig10}(d). We observe that the 
agreement between $\Delta I_{1}^{(r)(2)}$ and $\Delta I_{1}^{(r)(num)}$ 
is considerably better than the agreement between $\Delta I_{1}^{(r)(1)}$ 
and $\Delta I_{1}^{(r)(num)}$. In accordance with the latter observation, 
the value of the relative error $E_{r}^{(\Delta I_{1}^{(r)(2)})}(z_{f})$ 
is only $11.9\%$. Thus, based on the results shown in 
Figs. \ref{fig3}, \ref{fig4}, \ref{fig9}, and \ref{fig10} and similar 
results obtained with other values of the physical parameters, we conclude 
that our perturbation approach can indeed be used to design collision setups  
that lead to strong localized and nonlocalized intensity reduction effects.

% fig 10 the fractional intensity reduction factor vs x and y - setup 2 

\begin{figure}[ptb]
\begin{center}
\epsfxsize=12cm \epsffile{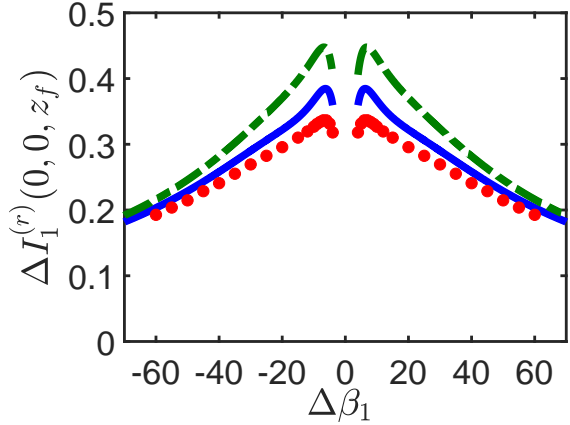}
\end{center}
\caption{(Color online) 
The fractional intensity reduction factor for pulsed-beam 1 at $(x,y)=(0,0)$ 
and $z=z_{f}$, $\Delta I_{1}^{(r)}(0,0,z_{f})$, vs the first-order dispersion 
coefficient $\Delta\beta_{1}$ in fast two-beam collisions with $\epsilon_{3}=0.25$ 
(collision setup 2). The red circles represent the result obtained by 
numerical simulations with Eq. (\ref{pb5}). The dashed-dotted green and 
solid blue curves represent the perturbation theory predictions 
of Eqs. (\ref{pb49}) and (\ref{pb77}), respectively.}  
\label{fig11}
\end{figure}

Since we are interested in the effects of cubic loss on fast collisions, 
it is important to study the dependence of the different collisional 
effects on $\Delta\beta_{1}$. We therefore carry out extensive numerical 
simulations with Eq. (\ref{pb5}) with the parameter values mentioned in 
the beginning of this subsection and with $\Delta\beta_{1}$ values in the 
intervals $4 \le |\Delta\beta_{1}| \le 60$. We measure the final value of 
the intensity reduction factor on the $z$ axis $\Delta I_{1}^{(r)}(0,0,z_{f})$ 
and the collision-induced amplitude shift $\Delta A_{1}^{(c)}$ as functions 
of $\Delta\beta_{1}$. Figure \ref{fig11} shows the $\Delta\beta_{1}$ 
dependence of $\Delta I_{1}^{(r)}(0,0,z_{f})$ obtained in the simulations  
together with the two perturbation theory predictions of 
Eqs. (\ref{pb49}) and (\ref{pb77}). Similar to the situation 
in setup 1, the values of $\Delta I_{1}^{(r)(num)}(0,0,z_{f})$ are larger 
than 0.19 over the entire $\Delta\beta_{1}$ intervals, in agreement with 
the perturbation theory predictions for strong intensity reduction.      
We also observe that $\Delta I_{1}^{(r)}(0,0,z_{f})$ has two local maxima 
at $\Delta\beta_{1} \approx \pm 7.0$, and that these local maxima are 
correctly captured by the two predictions of the perturbation theory. 
More generally, we find very good agreement between $\Delta I_{1}^{(r)(2)}(0,0,z_{f})$ 
and $\Delta I_{1}^{(r)(num)}(0,0,z_{f})$ and good agreement between 
$\Delta I_{1}^{(r)(1)}(0,0,z_{f})$ and $\Delta I_{1}^{(r)(num)}(0,0,z_{f})$ 
over the entire $\Delta\beta_{1}$ intervals. Furthermore, the agreement 
between the two theoretical predictions and the simulations result 
improves with increasing value of $|\Delta\beta_{1}|$. Therefore, 
based on the results seen in Fig. \ref{fig11}, we conclude that 
the significant intensity reduction effects in setup 2 are not 
limited to intermediate values of $|\Delta\beta_{1}|$, but are also 
observed for large $|\Delta\beta_{1}|$ values.

% fig 11 - the \Delta\beta_{1} dependence of the fractional intensity 
% reduction factor for pulsed-beam 1 at (x,y)=(0,0) and z=z_{f}. 
% collision setup 2  

\begin{figure}[ptb]
\begin{center}
\epsfxsize=12cm \epsffile{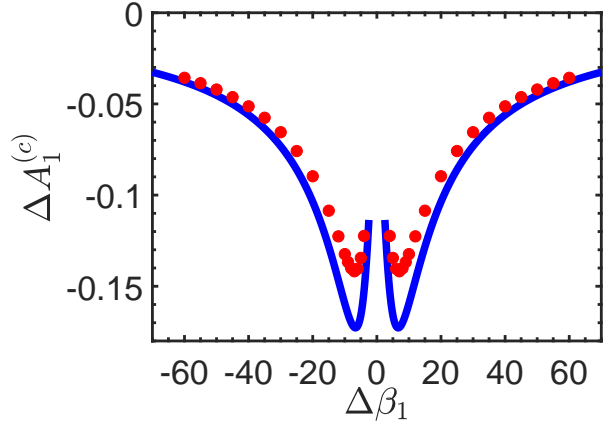}
\end{center}
\caption{(Color online) 
The collision-induced amplitude shift of pulsed-beam 1 $\Delta A_{1}^{(c)}$ 
vs the first-order dispersion coefficient $\Delta\beta_1$ in fast two-beam 
collisions with $\epsilon_3=0.25$ (collision setup 2). The red circles represent   
the result obtained by numerical simulations with Eq. (\ref{pb5}).  
The solid blue curve corresponds to the perturbation theory prediction  
of Eq. (\ref{pb71}).}  
\label{fig12}
\end{figure}

Figure \ref{fig12} shows the $\Delta\beta_{1}$ dependence of $\Delta A_{1}^{(c)}$  
that is obtained in the numerical simulations together with the perturbation 
theory prediction of Eq. (\ref{pb71}). We first observe that the values of 
$\Delta A_{1}^{(c)(num)}$ are of orders $10^{-2}-10^{-1}$, and are larger by 
one to two orders of magnitude compared with the $\Delta A_{1}^{(c)(num)}$ values in setup 1. 
This behavior can be explained by the difference in the nature of the intensity 
reduction in the two setups; the intensity reduction is nonlocalized in setup 2 
and localized in setup 1. As a result, the values of $\Delta A_{1}^{(c)}$, 
which are a measure for global intensity changes, are significantly larger in 
setup 2 compared with setup 1. We also observe good agreement between 
$\Delta A_{1}^{(c)(num)}$ and the theoretical prediction of Eq. (\ref{pb71}).     
In particular, the relative error in the approximation of $\Delta A_{1}^{(c)}$ 
is smaller than $19.7\%$ for $4 \le |\Delta\beta_1| <30$ and smaller than 
$10.5\%$ for $30\le |\Delta\beta_1| \le 60$. Thus, the dependence of 
$\Delta A_{1}^{(c)}$ on $\Delta\beta_{1}$ is correctly captured by our 
perturbation approach despite the strong collision-induced effects.

% fig 12 - \Delta A_{1}^{(c)} vs \Delta\beta_1 - setup 2 

Another important feature of the collision effects that is seen 
in Fig. \ref{fig12} is the existence of two local minima in the graph 
of $\Delta A_{1}^{(c)}$ vs $\Delta\beta_{1}$. The local minima are 
located at $\Delta\beta_{1} \approx \pm 7.0$, and are correctly captured 
by the perturbation theory prediction of Eq. (\ref{pb71}). 
In fact, using Eq. (\ref{pb71}) with $d_{21}=d_{22}=d_{2}$, we can show 
that $\Delta A_{1}^{(c)}$ has two local minima at $\Delta\beta_1 = \pm 2t_{20}d_{2}
(W_{10}^{(x)}W_{10}^{(y)}W_{20}^{(x)}W_{20}^{(y)})^{-1/2}.$         
Using the latter relation with the parameter values of setup 2, 
we obtain $\Delta\beta_{1} \approx \pm 6.67$, in good agreement with 
the simulations result. We emphasize that to our knowledge, this result  
represents the first observation of a deviation of the graph of  
$\Delta A_{1}^{(c)}$ vs $\Delta\beta_{1}$ from the common funnel 
shape that is observed in fast collisions between optical beams 
or optical pulses in the presence of weak nonlinear dissipation. 
Indeed, previous works on fast collisions between time-independent 
beams in spatial dimension 2 \cite{PHN2022}, and between temporal 
optical pulses in spatial dimension 1 \cite{PNH2017,NHP2022,QMN2022} in linear 
optical media with weak nonlinear loss showed that the graph of $\Delta A_{1}^{(c)}$ 
vs $\Delta\beta_{1}$ has a funnel shape and does not possess any local extrema. 
Similar results were obtained in Refs. \cite{PNC2010,PC2012,NH2021,NH2022} 
for fast collisions between optical solitons in the presence of 
weak nonlinear dissipation.         
Intuitively, the existence of the two local minima in Fig. \ref{fig12} 
is a result of a competition between two effects. On one hand, the factor 
$1/|\Delta\beta_1|$ in the first line of Eq. (\ref{pb71}), 
which is associated with the collision length $\Delta z_{c}$, 
leads to a decrease in the value of $\Delta A_{1}^{(c)}$ with increasing 
value of $|\Delta\beta_1|$. On the other hand, the 
dependence of the collision distance $z_{c}$ on $\Delta\beta_1$ 
in the expressions appearing in the second and third lines of 
Eq. (\ref{pb71}), which are associated with spatial beam spreading, 
leads to an increase in the value of $\Delta A_{1}^{(c)}$ with increasing 
value of $|\Delta\beta_1|$. The comparison between $\Delta A_{1}^{(c)(num)}$ 
and the prediction of Eq. (\ref{pb71}) shows that this delicate 
competition effect is also correctly captured by our perturbation approach.

\section{Conclusions}
\label{conclusions}
We studied fast collisions between two pulsed optical beams 
in a linear bulk optical medium with weak cubic loss, where the 
cubic loss is due to nondegenerate two-photon absorption. 
The collisions are induced by the difference between the 
first-order dispersion coefficients for the two pulsed-beams. 
These collisions are easier to realize than the collisions 
between the time-independent beams that were studied in 
Ref. \cite{PHN2022}. Indeed, the latter collisions are induced 
by beam-steering, and therefore, their realization requires the 
application of special techniques to control the beam-steering.  
We introduced a perturbation approach for calculating the 
effects of a single fast two-beam collision. The approach is 
based on the existence of two small parameters in the problem: 
the cubic loss coefficient $\epsilon_{3}$ and the reciprocal of 
the difference between the first-order dispersion coefficients 
$1/\Delta\beta_{1}$. We used the perturbation approach to obtain 
general formulas for the collision-induced changes in the 
pulsed-beam's shape and amplitude. Moreover, we used the 
approach to design and characterize collision setups that lead to strong 
localized and nonlocalized intensity reduction effects. More specifically, 
the design of these setups was based on the simple form of the expression 
for the relative intensity reduction factor in the collision interval, 
which was obtained by our perturbation approach. 
The values of the predicted collision-induced changes in the 
pulsed-beam's shape were larger by one to two orders of magnitude 
compared with the values obtained in fast collisions between 
time-independent beams in Ref. \cite{PHN2022} and in fast 
collisions between optical solitons in spatial dimension 1 
in Ref. \cite{PC2020}. The predictions of our perturbation theory 
were in good agreement with the results of numerical simulations 
with the perturbed linear propagation model for both localized and 
nonlocalized collision setups, despite the relatively strong 
collisional effects.

To gain further insight into the effects of cubic loss on fast 
two-beam collisions, we studied the dependence of the final value of the 
intensity reduction factor on the $z$ axis [$\Delta I_{1}^{(r)}(0,0,z_{f})$] 
and of the collision-induced amplitude shift ($\Delta A_{1}^{(c)}$) 
on $\Delta\beta_{1}$. We found that for both localized and nonlocalized  
collision setups, the values of $\Delta I_{1}^{(r)}(0,0,z_{f})$ were larger 
by two orders of magnitude compared with the values obtained in 
Ref. \cite{PHN2022} for fast collisions between time-independent beams. 
The predictions of the perturbation theory were in good agreement with 
the numerical simulations results over the entire $\Delta\beta_{1}$ intervals  
that were considered. Moreover, since the agreement between theory and 
simulations improved with increasing value of $|\Delta\beta_{1}|$, we 
concluded that the significant intensity reduction effects were not 
limited to intermediate $|\Delta\beta_{1}|$ values, but also existed 
for large $|\Delta\beta_{1}|$ values.

The $\Delta\beta_{1}$ dependence of the collision-induced amplitude 
shift that was found for the nonlocalized collision setups was different 
from the one found for the localized setups in both theory and 
simulations in two important aspects. First, the values of 
$\Delta A_{1}^{(c)}$ in the nonlocalized setups were larger by 
one to two orders of magnitude compared with the values obtained 
in the localized setups. This difference was explained by noting that 
$\Delta A_{1}^{(c)}$ is a measure for global intensity changes, 
and that as a result, the strong but localized intensity reduction 
in the localized setups leads only to relatively small amplitude shifts. 
Second, and more importantly, the graph of $\Delta A_{1}^{(c)}$ vs 
$\Delta\beta_{1}$ that was obtained for the nonlocalized setups contained 
two local minima at intermediate values of $\Delta\beta_{1}$. 
To our knowledge, this finding represents the first observation of a 
deviation of the graph of $\Delta A_{1}^{(c)}$ vs $\Delta\beta_{1}$ 
from the common funnel shape that is obtained in fast collisions 
between temporal pulses or time-independent beams in linear optical media 
\cite{PNH2017,NHP2022,PHN2022,QMN2022}, and in fast collisions between 
optical solitons \cite{PNC2010,PC2012,NH2021,NH2022}. 
The existence of the local minima in the graph of $\Delta A_{1}^{(c)}$ vs 
$\Delta\beta_{1}$ was correctly captured by our perturbation theory. 
It was explained as a result of a competition between two effects that 
depend on $\Delta\beta_{1}$ and that affect the magnitude of 
$\Delta A_{1}^{(c)}$ in opposite manners.

In summary, we presented the first study of fast collisions between 
optical pulses (or optical beams) in the presence of weak nonlinear loss, 
in which the observed collision-induced effects are strong. 
We also characterized for the first time the differences  
between localized and nonlocalized effects in these collisions. 
Our perturbation approach played an important role in the design 
of the collision setups, and in the characterization and analysis 
of the strong collisional effects. In this manner, our work 
significantly extended the results of previous studies on fast 
collisions between optical pulses or time-independent beams in  
linear media with weak nonlinear loss \cite{PNH2017,NHP2022,PHN2022,QMN2022}, 
and on fast collisions between optical solitons in the presence of 
weak nonlinear loss \cite{PNC2010,PC2012,NH2021,NH2022,PC2020}. 
Indeed, all these previous works on the subject were limited to 
weak collision-induced effects, and did not characterize the 
differences between localized and nonlocalized effects. 
In view of the strong collisional effects that we observed and the 
fact that our perturbative calculation is based on a number of simplifying 
assumptions, whose validity conditions are not known, the good 
agreement between the perturbation theory and the simulations 
in the current work is quite surprising. The relatively strong 
intensity reduction effects can be very useful for spatial reshaping 
of pulsed optical beams. Additionally, our results can be useful for 
WDM (multisequence) optical communication systems due to the importance 
of fast collisions between optical pulses and optical beams in these systems.

%\section*{Acknowledgments}

\appendix
\section{Derivation of Eq. (\ref{pb68})}
\label{appendA}   
   
In this Appendix, we derive Eq. (\ref{pb68}) for the inverse 
Fourier transforms of $\hat g_{12}^{(x)}(k_{1},z_{c})
\exp[-i d_{21} k_{1}^{2}(z-z_{c})]$ and 
$\hat g_{12}^{(y)}(k_{2},z_{c}) \exp[-i d_{21} k_{2}^{2}(z-z_{c})]$ 
in the case where the initial condition is given by Eq. (\ref{pb45}).      
The derivations of the expressions for the two inverse Fourier 
transforms are very similar. In fact, we can derive the two expressions 
at the same time and express the end results in a single formula 
[Eq. (\ref{pb68})]. For this purpose, we introduce the symbol $u$, 
which stands for the indexes 1 or 2 in $k_{u}$, and for $x$ or $y$, 
respectively, everywhere else in the derivations. With this notation, 
we only need to obtain a single expression for the inverse Fourier transform 
of $\hat g_{12}^{(u)}(k_{u},z_{c})\exp[-i d_{21} k_{u}^{2}(z-z_{c})]$.

Using the expressions for the solutions of the unperturbed linear 
propagation equation, we obtain: 
\begin{eqnarray}&&
g_{12}^{(u)}(u,z_{c})=
\frac{W_{10}^{(u)}W_{20}^{(u)2}}
{(W_{10}^{(u)4} + 4d_{21}^2 z_{c}^{2})^{1/4} 
(W_{20}^{(u)4} + 4d_{22}^2 z_{c}^{2})^{1/2}}
\nonumber \\&&
\times 
\exp\left[-\tilde a_{2}^{(u)2}(z_{c})u^{2}
-\frac{i}{2}\arctan\left(\frac{2d_{21}z_{c}}{W_{10}^{(u)2}}\right)\right], 
\label{appA1}     
\end{eqnarray}  
where 
\begin{equation}
\tilde a_{2}^{(u)2}(z_{c})= q_{1}^{(u)}(z_{c})+ iq_{2}^{(u)}(z_{c}), 
\label{appA2}
\end{equation}    
\begin{eqnarray}&&
\!\!\!\!\!\!\!\!\!\!\!
q_{1}^{(u)}(z_c)=
\frac{W_{10}^{(u)2}}{2(W_{10}^{(u)4} + 4d_{21}^{2}z_{c}^{2})}
+\frac{W_{20}^{(u)2}}{W_{20}^{(u)4} + 4d_{22}^{2}z_{c}^{2}}, 
\label{appA3}     
\end{eqnarray}  
and 
\begin{eqnarray}&&
q_{2}^{(u)}(z_c)=\frac{-d_{21}z_{c}}{W_{10}^{(u)4} + 4d_{21}^{2}z_{c}^{2}} .
\label{appA4}     
\end{eqnarray}     
The Fourier transform of $g_{12}^{(u)}(u,z_{c})$ is 
\begin{eqnarray}&&
\hat g_{12}^{(u)}(k_{u},z_{c})=
\frac{W_{10}^{(u)}W_{20}^{(u)2}(W_{10}^{(u)4} + 4d_{21}^{2}z_{c}^{2})^{1/4}}
{\tilde a_{3}^{(u)}(z_{c})}
\nonumber \\&&
\times 
\exp\left[-\frac{k_{u}^{2}}{4\tilde a_{2}^{(u)2}(z_{c})}
-\frac{i}{2}\arctan\left(\frac{2d_{21}z_{c}}{W_{10}^{(u)2}}\right)\right], 
\label{appA5}    
\end{eqnarray}   
where  
\begin{equation}
\tilde a_{3}^{(u)}(z_{c})=
\left[2(W_{10}^{(u)4} + 4d_{21}^{2}z_{c}^{2})
(W_{20}^{(u)4} + 4d_{22}^{2}z_{c}^{2})\right]^{1/2}
\tilde a_{2}^{(u)}(z_{c}).  
\label{appA6}
\end{equation}  
Using Eqs. (\ref{appA2})-(\ref{appA6}), we find that the inverse Fourier 
transform of \linebreak 
$\hat g_{12}^{(u)}(k_{u},z_{c})\exp[-i d_{21} k_{u}^{2}(z-z_{c})]$ 
is given by:  
\begin{eqnarray}&&
{\cal F}^{-1}\left(\hat g_{12}^{(u)}(k_u,z_c)
\exp[-id_{21} k_u^2(z-z_c)]\right) =
\nonumber\\&&
=\frac{W_{10}^{(u)}W_{20}^{(u)2} 
\exp\left[-q_{1}^{(u)}(z_c)u^2/R_{1}^{(u)4}(z,z_c) 
+ i\chi_1^{(u)}(u,z)\right]}
{(W_{10}^{(u)4} + 4d_{21}^2 z_c^2)^{1/4} 
(W_{20}^{(u)4} + 4d_{22}^2 z_c^2)^{1/2} R_{1}^{(u)}(z,z_c)},
\label{appA7}
\end{eqnarray} 
where 
\begin{eqnarray}&&
\!\!\!\!\!\!\!\!\!\!\!\!\!\!\!\!\!\!\!\!\!
R_{1}^{(u)}(z,z_c) = \left\{1 - 8d_{21}q_{2}^{(u)}(z_c) (z-z_{c}) 
%\right.
%\nonumber\\&&
%\left.
+ 16d_{21}^{2}\left[q_{1}^{(u)2}(z_c) 
+ q_{2}^{(u)2}(z_c)\right](z-z_{c})^{2}\right\}^{1/4},
\label{appA8}
\end{eqnarray}  
and 
\begin{eqnarray}&&
%\!\!\!\!\!\!\!\!\!\!\!\!\!\!\!\!\!\!\!\!\!
\chi_{1}^{(u)}(u,z) = -\frac{1}{2}\arctan\left(\frac{2d_{21}z_{c}}{W_{10}^{(u)2}}\right) 
-\frac{1}{2}\arctan\left[\frac{4d_{21}q_{1}^{(u)}(z_c) (z-z_{c})}
{1 - 4d_{21}q_{2}^{(u)}(z_c) (z-z_{c})}\right]
\nonumber\\&&
- \left[q_{2}^{(u)}(z_c) - 4d_{21}\left(q_{1}^{(u)2}(z_c) 
+ q_{2}^{(u)2}(z_c)\right)(z-z_{c})\right]
\frac{u^{2}}{R_{1}^{(u)4}(z,z_c)} .
\label{appA9}
\end{eqnarray}  
Equation (\ref{appA7}) is Eq. (\ref{pb68}) of 
subsection \ref{beam_shape}.

%\section*{References}
  

\begin{thebibliography}{}
\bibitem{fast_collisions} In particular, the collision length is 
much smaller than the dispersion length and/or the diffraction length.
\bibitem{parameters1} Notice that with this choice, 
$\beta_{22}=1.0060$, $d_{21}=0.06$, and $d_{22}=0.06009$. 
However, since the values of the collision-induced changes in the 
shapes and amplitudes of the pulsed-beams are not sensitive 
to the exact values of these parameters, we use for simplicity 
$\beta_{22}=1.0$ and $d_{22}=0.06$ in the numerical simulations in setup 1.
\bibitem{cross_sections} We choose to present the pulsed-beam shapes 
using cross-sections, since this enables a clearer presentation of the 
dynamics of the main bodies of the pulsed-beams compared with 
conventional contour plots. Indeed, in the latter plots,  
the main bodies of the pulsed-beams are typically obscured by 
the outer parts (i.e., by the pulsed-beams tails). 
The three cross-sections are chosen as the planes, 
$x=0$, $t=0$, and $t=t_{2}(z)$, where $t_{2}(z)$ is the 
time coordinate of the center of pulsed-beam 2 at distance $z$.
\bibitem{zi_values} The value of $z_{i}$ is determined by:  
$z_{i}=z_{c}+r(z_{f}-z_{c})$, where $r=1/5$, as an example. 
Thus, $z_{i}$ is an intermediate distance that is larger 
than $z_{c}$, at which the collision is not yet completed.
\bibitem{parameters2} Notice that with this choice, 
$\beta_{22}=1.0060$, $d_{21}=1.0$, and $d_{22}=1.0014$. 
However, since the values of the collision-induced changes in the 
pulsed-beams' shapes and amplitudes are not sensitive 
to the exact values of these parameters, we use for simplicity 
$\beta_{22}=1.0$ and $d_{22}=1.0$ in the numerical simulations in setup 2.
\end{thebibliography}
\end{document}